\documentclass[traditabstract]{aa} 

\newcommand{\coralie}{{\tiny CORALIE}}
\newcommand{\sophie}{{\tiny SOPHIE}}
\newcommand{\elodie}{{\tiny ELODIE}}
\newcommand{\narval}{{\tiny NARVAL}}
\newcommand{\harps}{{\tiny HARPS}}

\newcommand{\kms}{km~s$^{-1}$}
\newcommand{\ms}{m~s$^{-1}$}

\usepackage{graphicx,amssymb,amsmath,natbib,hyperref,multirow}
\usepackage[usenames,dvipsnames]{xcolor}
\usepackage{rotating}
\usepackage{txfonts}

\definecolor{mygreen}{rgb}{0.13,0.73,0.13}

\begin{document}
\title{The catalogue of radial velocity standard stars for Gaia 
\thanks{Based on data obtained within the Gaia DPAC (Data Processing and
Analysis Consortium) and coordinated by the GBOG (Ground-Based
Observations for Gaia) working group, at various telescopes; see abstract}
\thanks{Tables are only available in electronic form at the CDS via anonymous ftp to cdsarc.u-strasbg.fr (130.79.128.5) 
or via http://cdsarc.u-strasbg.fr/viz-bin/qcat?J/A+A/552/A64}
}

\subtitle{I. Pre-launch release}

 \author{C. Soubiran\inst{1}
 \and G. Jasniewicz\inst{2}
 \and L. Chemin\inst{1}
 \and F. Crifo\inst{3}   
 \and S. Udry\inst{4}   
 \and D. Hestroffer\inst{5}   
 \and D. Katz\inst{3}   }

\institute{LAB UMR 5804, Univ. Bordeaux  - CNRS, F-33270, Floirac, France\\
             \email{soubiran@obs.u-bordeaux1.fr}
\and LUPM UMR 5299 CNRS/UM2, Universit\'e Montpellier II, CC 72, 34095 Montpellier Cedex 05, France
\and GEPI, Observatoire de Paris, CNRS, Universit\'e Paris Diderot, 5 place Jules Janssen, 92190 Meudon, France
\and Observatoire de Gen\`eve, Universit\'e de Gen\`eve, 51 Ch. des Maillettes, 1290 Sauverny, Switzerland
\and  IMCCE, Observatoire de Paris, UPMC, CNRS UMR8028, 77 Av. Denfert-Rochereau, 75014 Paris, France}

   \date{Received \today; accepted XX}

  \abstract{
  The Radial Velocity Spectrograph (RVS) on board of Gaia needs to be calibrated using stable
   reference stars known in advance. The catalogue presented here was built for that purpose. It includes 
   1420 radial velocity standard star candidates selected on strict criteria to fulfil 
   the Gaia-RVS requirements.
  A large programme of ground-based observations has been underway since 2006 to monitor these stars 
  and verify their stability, which has to be better than 300 \ms\ over several years.
  The observations were done on the echelle spectrographs \elodie\ and \sophie\ on the 1.93-m 
  telescope at Observatoire de Haute-Provence (OHP),  \narval\ on the T\'elescope Bernard Lyot at
   Observatoire du Pic du Midi and \coralie\ on the Euler-Swiss Telescope at La Silla.  
   Data from the OHP and  Geneva Observatory archives have also been retrieved as have \harps\ spectra from the ESO archive.  
   We provide a mean radial velocity in the \sophie\ scale for each star, derived from the combination of velocities 
   measured with those instruments, 
   after having carefully estimated their differences in zero points. In total, 10214 radial velocity 
   measurements have been obtained for the 1420 stars. With a mean time baseline of 6.35 years, 92.9\% of the candidates fulfil  a target  stability  
  criterion of 300 \ms. Three hundred forty-three stars are found to be constant at the level of 100 \ms\ over 10 years. Comparisons  with  earlier catalogues show  excellent agreement for FGK stars, with zero-point differences lower than 100 \ms\ and a remarkably low RMS scatter of 33 \ms\ in one case, suggesting that the precision of the catalogue presented here is better than this value. This catalogue will likely be useful for other large-scale
spectroscopic surveys, such as APOGEE, Gaia-ESO, HERMES, and LAMOST.}



   \maketitle
%

\section{Introduction}
\label{sec:intro}
Gaia, the next astrophysical mission of the European Space Agency (ESA), will be launched in 
2013\footnote{{http://www.rssd.esa.int/index.php?project=GAIA\&page=index}}. It will survey the entire sky during five years, and measure the positions, proper motions  and parallaxes of some $10^9$ stars down to a visual magnitude of  20, with an expected parallax accuracy of 10 to 25 $\mu$as at 15$^{\rm th}$ magnitude \citep{deb12}. Stars classified as single will have their effective temperature, surface gravity, and metallicity estimated, as well as  interstellar extinction. 

The Radial Velocity Spectrometer (RVS) on board Gaia will provide radial velocities (RV) of about 150 million stars down to  17$^{\rm th}$ magnitude with precisions ranging from 15 \kms\  at the faint end to 1 \kms\ or  better for G and K stars at magnitudes brighter than $V \simeq 13.5$  \citep{kat04,kat09}. 
 The RV, combined with astrometry, will give access to the six dimensions of the phase space useful for kinematical studies. The RVS will also provide rotational velocities and atmospheric parameters for about five million stars down to $V \simeq 13$ and iron and $\alpha$-elements abundances for about two million stars down to $V \simeq 12$. 
 Such a large spectroscopic survey will have a tremendous impact on many science cases, 
 such as the chemistry and dynamics of the Milky Way, 
 the detection and characterization of multiple systems, and variable stars. 
 More details of the expected  science yield from the RVS are described in \citet{wil05}.

The RVS has a resolving power of 11500 and covers the spectral range $847-874$ nm, which includes the near-infrared Ca{\sc ii} triplet, 
many lines of iron and $\alpha-$elements, and Paschen lines in hot stars. 
This spectral range has already proven to be very well suited to RV analysis with the RAVE survey \citep{ste06}. 
The RVS is an integral field spectrograph with no entrance slit and no
on-board wavelength calibration source. As a consequence, the RVS wavelength dispersion law will be
derived from the RVS observations. The instrument will be
self calibrated. As explained in \citet{jas11}, ground-based radial velocity standards are
mandatory for calibrating the wavelength zero point, which otherwise
would be mathematically degenerated with the radial velocity scale
(i.e. a shift in the wavelength scale can be compensated for by a shift
in the RV scale). The Gaia Data Processing and Analysis Consortium (DPAC), and more specifically its Development Unit 'Spectrocopic Processing' \citep{kat11}, is responsible for establishing the requirements for this external calibration and for the acquisition of the necessary auxiliary data

In principle asteroids are excellent sources for calibrating the zero point because
their radial velocity can be derived from celestial mechanics with
uncertainties lower than 1 \ms. However,  they cannot be the main reference sources for Gaia because 
they are not numerous enough in the appropriate magnitude range and their sky distribution is limited to the Ecliptic Plane. 

When we started to search for suitable RV standard stars six years ago, there was no existing dataset fulfilling 
 the Gaia RVS requirements in terms of sky coverage, magnitude range, number of targets, and stability. 
The only official list of standard stars was the one compiled by the IAU Commission 30, which did not turn out to be suitable for our purpose. 
We thus built a stellar grid consisting of 1420  FGKM stars in
the magnitude range $6<V<11$, with a homogeneous distribution on the sky. The selection process of  that sample is 
fully explained in \citet{cri10}. We briefly summarize it in Section~\ref{sec:candidates}. 

The RVS calibration requires standard stars with much better RV stability than the 1 \kms\ accuracy expected from their future  measurements with RVS,  with no drift until the end of the mission (2018). The value of 300 \ms\ is adopted as the stability level to be checked.
To qualify as a reference star, each 
candidate  has to be observed at least  twice before launch and another time during the mission  
to verify its long-term stability. The RV measurements available to date are described 
in Section~\ref{sec:measurements}, and their combination into a homogeneous scale in Section~\ref{sec:combination}. We compare this new 
catalogue to other studies in  Section~\ref{sec:comparison}. The stablest stars are presented in Sect.~\ref{s:stable_stars} as are considerations about variable stars. Future work is described in Sect.~\ref{s:next}.

\section{Radial velocity standard-star candidates}
\label{sec:candidates}
The selection of stars suitable for building the reference grid for RVS has been carefully
considered and several criteria defined \citep{cri10}. The list of candidates was 
established from three catalogues: ``Radial velocities of 889 late-type stars'' \citep{nid02}, 
``Radial velocities for 6691 K and M giants'' \citep{fam05}, and ``The Geneva-Copenhagen Survey 
of Solar neighbourhood'' \citep{nor04}, complemented with IAU standards \citep{udr99}. 
We selected the stars  having the best observational history in terms of 
 consistency of radial velocity measurements over several years. 
In that way we could focus on stars  more likely to  be stable over time. 
Moreover, since all the candidates are in the HIPPARCOS Catalogue we were able to check their properties
(photometry, spectral type, variability, multiplicity, etc.) in a homegeneous way.  
Another important criterion was that each candidate had to have no close neighbour within a circular region of 80\arcsec-radius.   
This criterion was set to ensure that their future RVS spectra will not be contaminated by another overlapping spectrum, 80\arcsec\ being the length of an RVS spectrum projected on the sky. 
This criterion was verified thanks to the USNO-B1 catalogue. 
These different criteria led to a preliminary list of 1420 candidates having a high probability of being RV-stable and well suited to becoming reference stars for the RVS.

\begin{figure}[htp]
\begin{center}
 \includegraphics[width=\columnwidth]{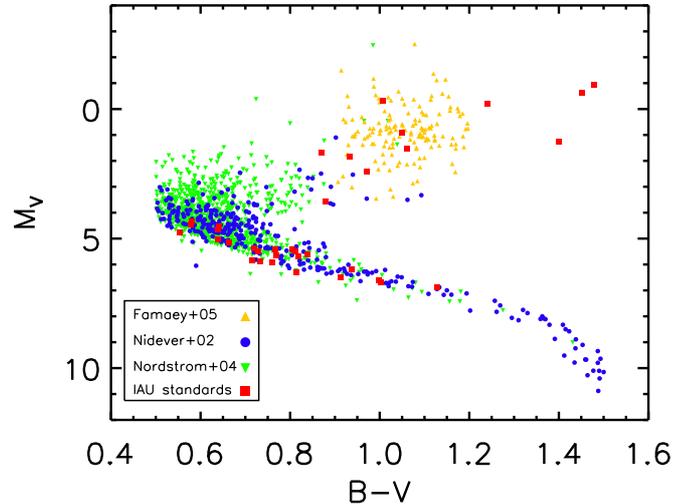}       
 \caption{HR diagram of the 1420 selected stars, with  a colour code corresponding to their origin.}
\label{f:HR}
\end{center}
\end{figure}

All of the 1420 candidates have at least three previous RV measurements in the above-mentioned catalogues; 
however, since they have been obtained during observational runs spanning irregular time baselines,
 their stability has not yet been ensured, particularly not until the end of the Gaia mission. 
 The main cause of RV variability at the level of 300 \ms\ and above is binarity, but the risk of having binary stars with significant amplitude in our sample has been reduced 
 because of all these selection criteria  and a careful analysis of the RV measurements available  for them so far.
 The HR diagram of the 1420 candidates is shown in Fig.~\ref{f:HR}, as originally published in \citet{cri10}.

\section{Radial velocity measurements}
\label{sec:measurements}
 The RV standards to be used for the RVS calibrations should not exhibit variations larger than 300 \ms\ during the five years of the Gaia observations, between 2013 and 2018. Since the standards need to be known before launch, we extrapolated this criterion to the current period in order to perform a pre-qualification of the candidates according to their stability during the time span of our ground-based observations, between 1995 and 2012. 
Considering that the candidates already have a good observational history, 
our strategy of pre-qualifying them is (i) to obtain at least two
new measurements per star before the Gaia launch,  separated   by more than one year, and (ii) to verify that the variation in these new measurements does not exceed 300 \ms. Then supplementary ground-based observations will have to be done during the mission to verify the long-term stability of the reference grid. 

Since the candidates are distributed all over the sky, observations in both
the northern and southern hemispheres  need to be performed with different instruments. An observing plan has been prepared under the umbrella of the GBOG (Ground Based Observations for Gaia) working group in charge of acquiring auxiliary data for the Gaia processing.  Eighty nights have been obtained over six years for that programme on four high-resolution spectrographs:  
\elodie\ then \sophie\ mounted on the 1.93-m telescope at Observatoire de Haute-Provence (OHP), 
 \narval\ on the T\'elescope Bernard Lyot at Observatoire du Pic du Midi, and \coralie\ on the Euler-Swiss telescope at La Silla. 

 RV measurements were also retrieved from the public archives of \sophie\ and \elodie\ 
at OHP \citep{mou04}, and of \harps\ at the ESO Advanced Data Products database. Many measurements are available in these archives, in particular due to the various exoplanet-hunting programmes.
 The full \elodie\ archive was not public when the observations started in 2006, and the volume of the other archives increased dramatically during the course of our programme. The recent query of the archives allowed us to retrieve several thousands of relevant measurements. There were many time series available for some stars. For time series in one single night, only the observation with the highest signal-to-noise ratio was considered. Additional data from the private \coralie\ archive have also been provided by the Geneva Observatory.

\sophie, \coralie, \elodie\ and \harps\ have respectively resolving powers of 
$R=\lambda/\Delta\lambda,\sim 75\,000, 50\,000, 42\,000$, and $120\,000$, respectively, and cover the visible range. 
They all have similar automatic on-line data reduction software to derive the barycentric RV by cross-correlation 
of  the spectra with a numerical mask \citep{bar96}. The spectral type of the numerical mask is chosen to 
be the closest to the  observed star.  It can be G2, K5, M4, and M5 for \sophie, G2, K5 and M2 for \harps\ and \coralie, F0 and K0 for \elodie.

\narval\ has a resolving power of $R \sim 78\,000$. Since this intrument has not been built for 
RV measurements but for polarimetry, the on-line reduction software does not include the RV determination. We measured it 
by cross-correlating the observed spectra with the \sophie\ G2 mask. Our main interest in observing with \narval\ is that 
its spectral coverage includes the RVS range (847$-$874 nm), allowing us to investigate possible systematic differences occurring when 
measuring the RV in the full visible range and in that narrow NIR spectral range. 
It is, however, beyond the scope of this paper to enter into these considerations.

Errors on radial velocities come from the photon noise, the wavelength calibration and some possible instrumental or 
other effects, which are impossible to quantify. It is, however, important to estimate errors, noted $\epsilon$ in the following, for the weighting scheme  adopted to combine measurements taken with different instruments and under various observing conditions.  The photon-noise uncertainty is used, deduced from the width and contrast of the cross-correlation function (hereafter CCF) and from the signal-to-noise 
ratio of the spectrum \citep{bar96, bou05}, and provided by  the data reduction software. 
External systematic errors are then quadratically added to the photon-noise uncertainty: 0.8 \ms\ for \harps\ (F. Bouchy, private communication), 4 \ms\ for \sophie\ \citep{boi12}, 5 \ms\ for \coralie\ (D. Naef, private communication), and 15 \ms\ for \elodie\ \citep{bar96}.  These systematic errors have been estimated for the most precise observing mode of the instruments where the stellar spectrum is recorded simultaneously with a thorium-argon calibration. But not all the \harps, \sophie, and \elodie\ spectra in the sample presented here have been obtained in that mode, and the corresponding RV measurements have larger errors. This is taken into account by doubling the systematic error listed above for exposures without the simultaneous calibration. It is worth recalling that $\epsilon$ estimated that way does not represent the total RV uncertainty, but is only meant to combine relative measurements with proper weights.

Table~\ref{t:nb_mes} summarizes the status of the observations, providing the number of spectroscopic measurements obtained so far with each instrument (N$_{\rm RV}$), the corresponding number of stars (N$_*$), the average errors of individual RV measurements $\overline\epsilon$, and the date range of the observations. In total there are 10214 RV measurements.

\begin{table}[h]
  \centering 
  \caption{Summary of RV measurements available as of November 2012, average of individual RV errors, and epoch range of the measurements.}
  \label{t:nb_mes}
\begin{tabular}{l|c|c|c|c}
\hline
             & N$_{\rm RV}$ &  N$_*$ & $\overline \epsilon$  & date  \\
             &  &  &  \ms & range \\
\hline
  \sophie &  2945   &  729   &  6.7   & 2006 - 2012 \\
\hline
  \coralie &  2470   &  775   &  12.9  & 1999 - 2012  \\
\hline
 \narval &  213   &  157   &  17.6   & 2007 - 2012 \\
\hline
 \elodie &  3931   &   372  & 17.4   & 1995 - 2006 \\
\hline
 \harps &  655   &  113   &   2.2 & 2003 - 2009 \\
\hline
\end{tabular}
\end{table}

\begin{figure*}[ht!]
\begin{center}
 \includegraphics[width=14cm]{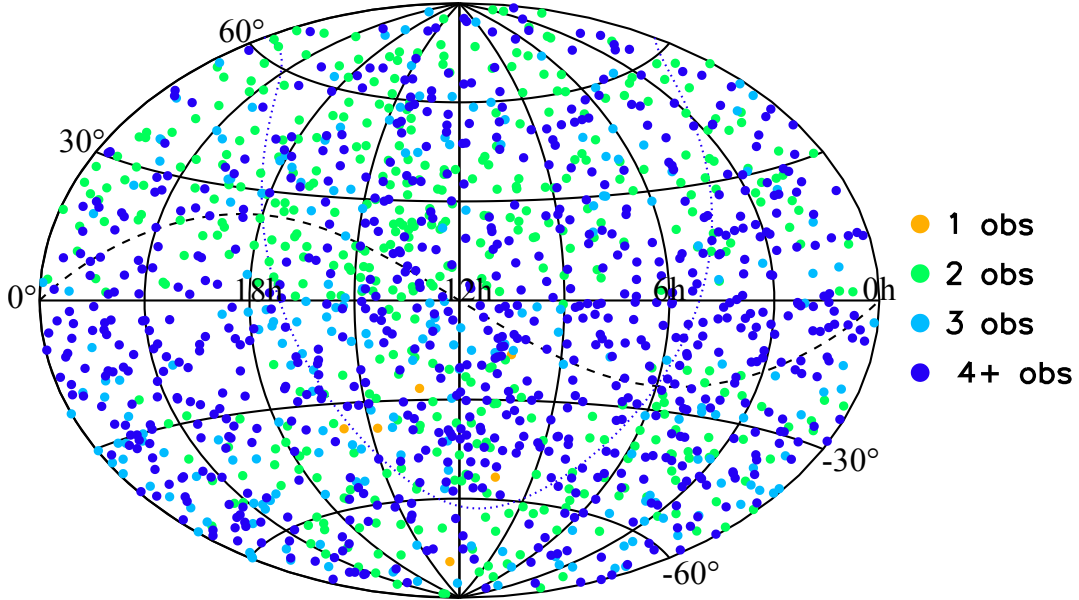}      
 \caption{Distribution of the 1420 stars on the celestial sphere in equatorial coordinates. The colour code indicates the number of measurements 
 obtained for each star. A dashed line indicates the projection of the Ecliptic plane, a dotted line that of the Galactic plane.}
\label{f:sky}
\end{center}
\end{figure*}

Figure~\ref{f:sky} represents the distribution of the 1420 candidate stars on the celestial sphere,
 with a colour code indicating the number of measurements  obtained for each star. As expected, at least two measurements are available for the vast majority of the stars. One measurement is missing for only a few stars in the southern hemisphere.

\begin{figure}[ht!]
\begin{center}
 \includegraphics[width=\columnwidth]{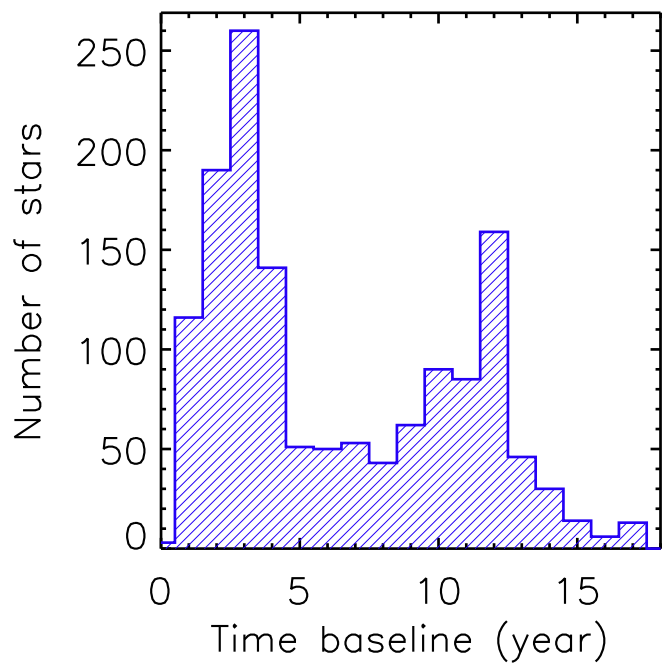}      
 \caption{Distribution of time baselines of observations for the 1420 candidates.}.
\label{f:histo_time}
\end{center}
\end{figure}

Figure~\ref{f:histo_time} shows the histogram of the time span between the first and last measurements of each of the 1420 
candidates. This is 6.35 years on average (median 4.37 years), up to 17 years thanks to the \elodie\ archive.  

\section{Combining RVs from different instruments}
\label{sec:combination}
An important step in the process of determining whether a star has a stable radial velocity or not is to combine  the measurements obtained with the 
 different  spectrographs. This requires taking the individual zero point of each instrument into account. 
 The zero point is related to the instrument itself, its spectral range, and resolution, and to the calibration procedure and 
  the method used to derive an RV. 
  In our case the same algorithm of cross-correlation with a numerical mask has been used that 
  should minimize the offsets between instruments. 
However, since  different masks have sometimes been used, a dependency of the offsets on the colour is expected. 
We adopted the \sophie\  frame as the radial velocity reference scale for all spectrographs, and we made certain to have a sufficient number of stars observed in common with \sophie\  and with the other instruments to derive the offsets. 

To measure the offset between \sophie\  and the other instruments, 
we first computed the weighted mean and standard deviation  of the RV measurements
 per instrument and per star. The weight of each individual measurement RV$_i$ is 
 defined as $1/\epsilon_i^2$, $\epsilon_i$ being the RV error mentioned in Sect.~\ref{sec:measurements}.

\begin{figure*}[h!]
\begin{center}
 \includegraphics[width=0.4\textwidth]{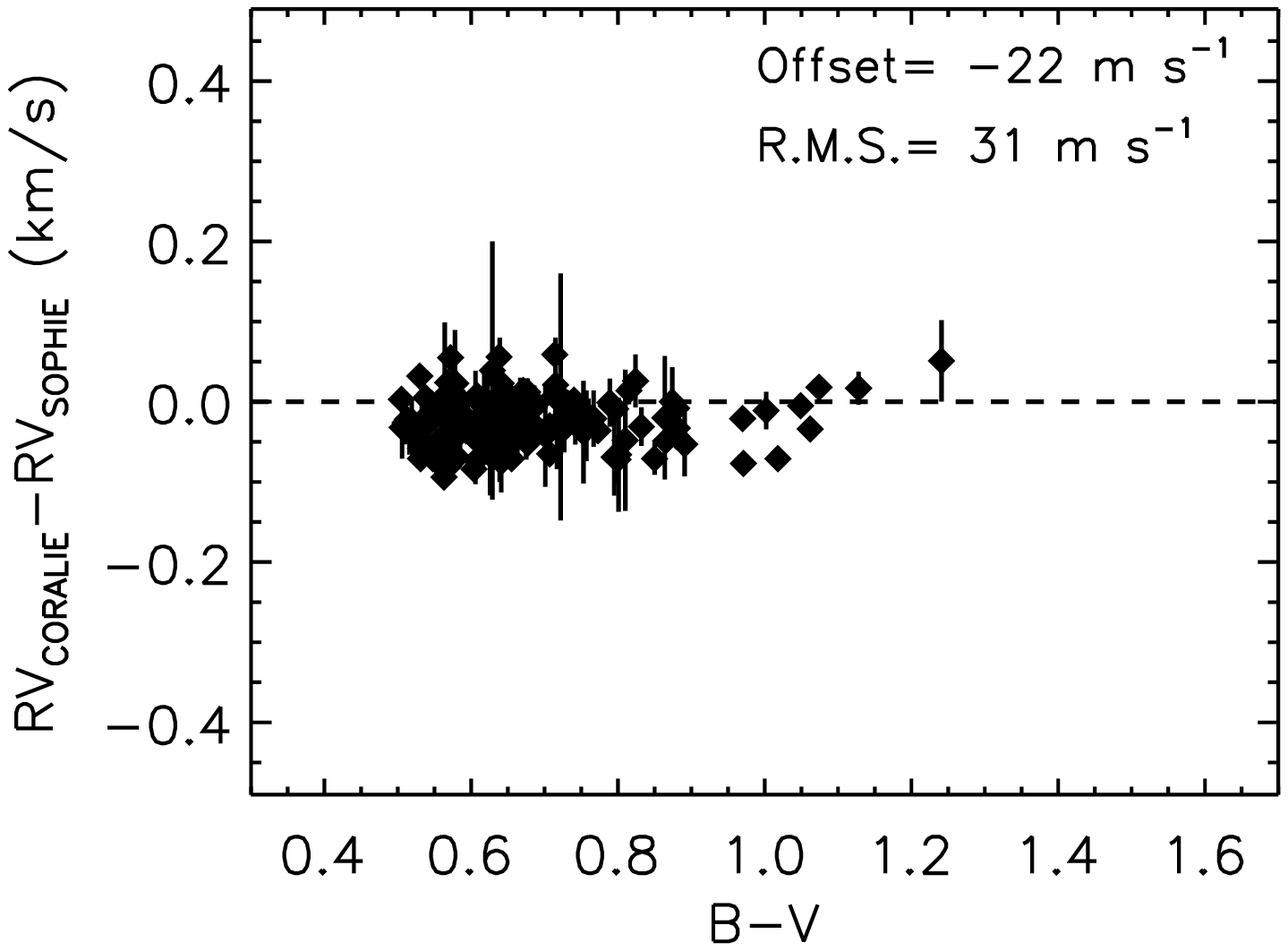}\includegraphics[width=0.4\textwidth]{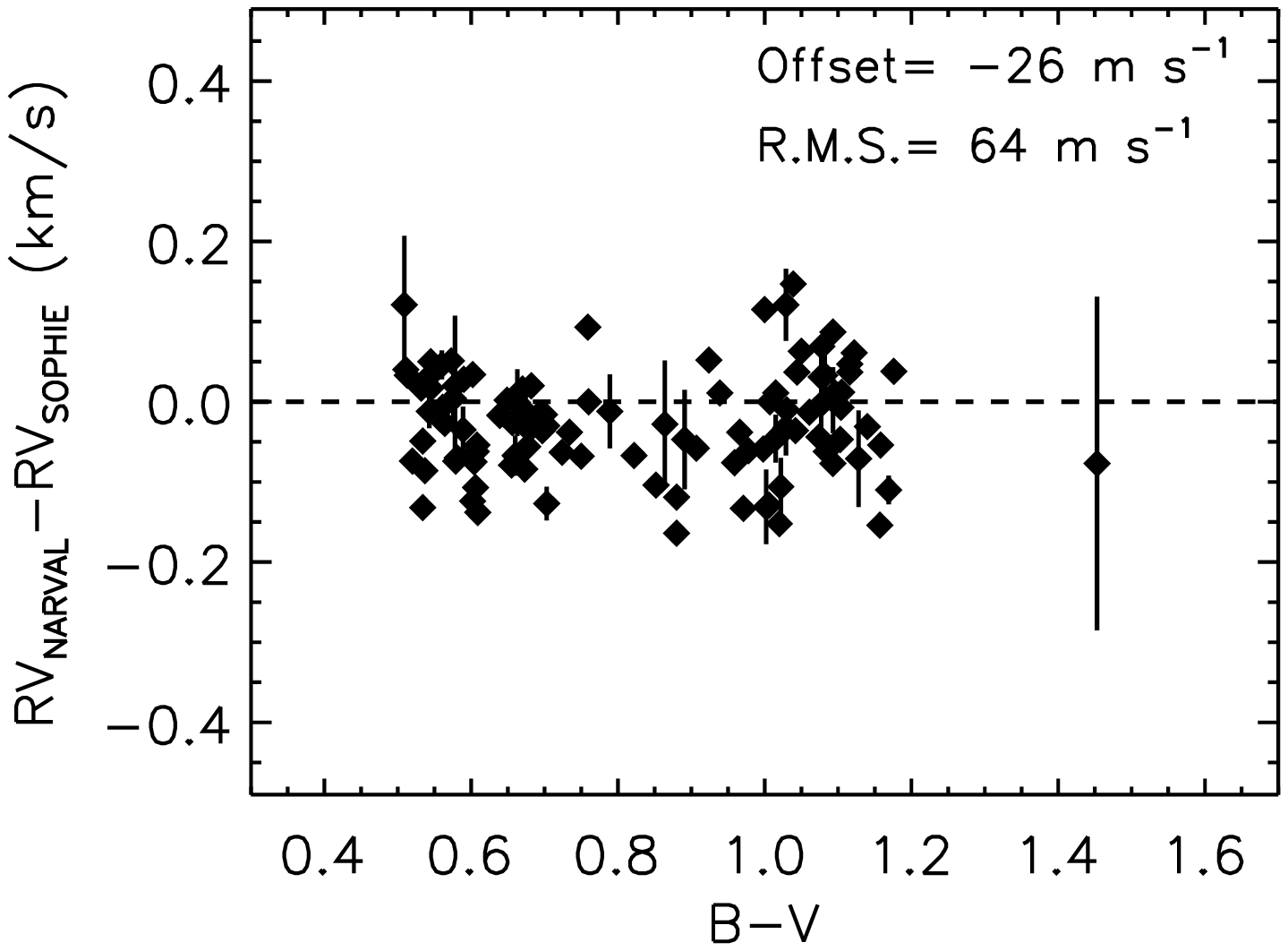}      
 \includegraphics[width=0.4\textwidth]{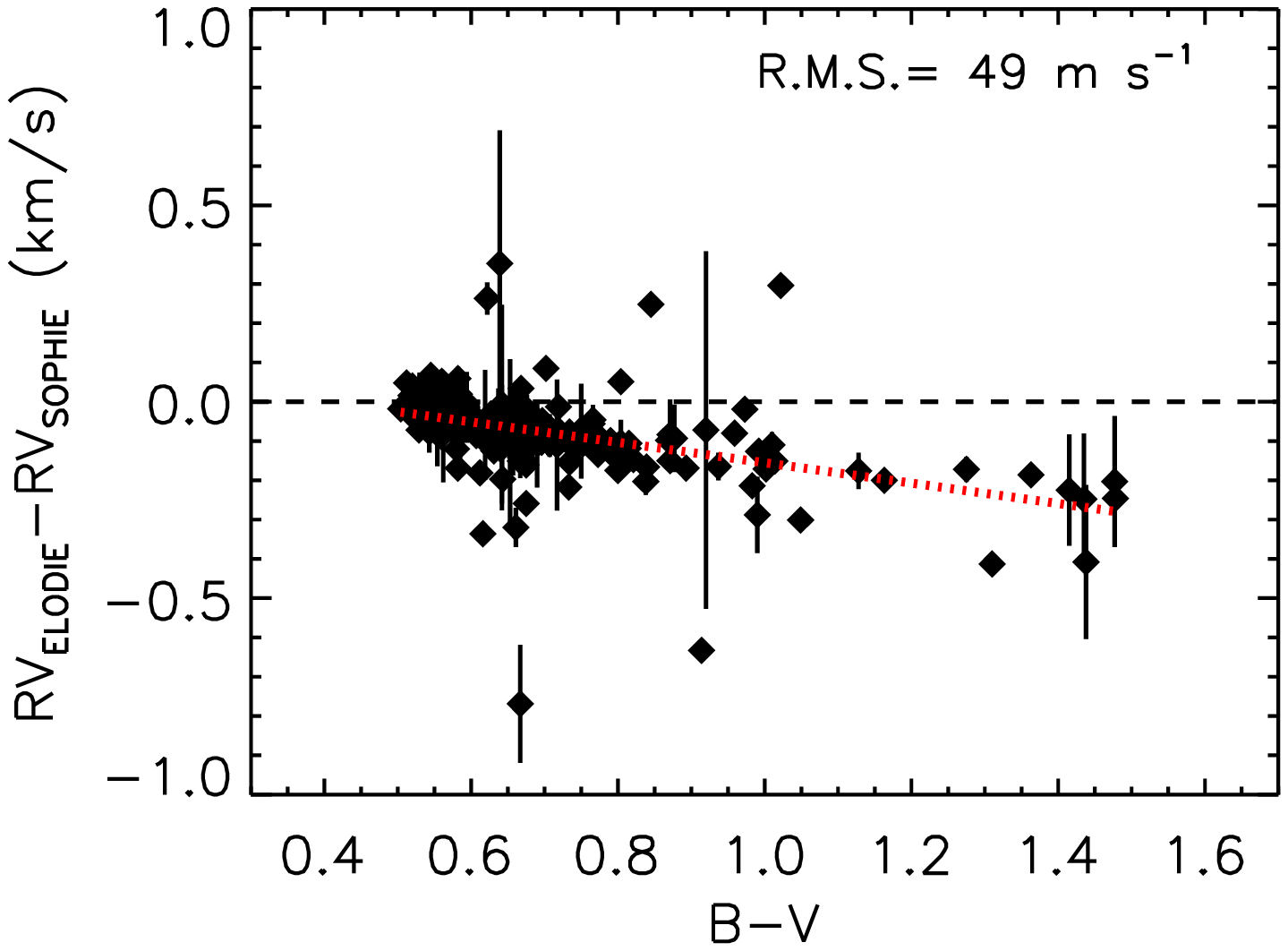}\includegraphics[width=0.4\textwidth]{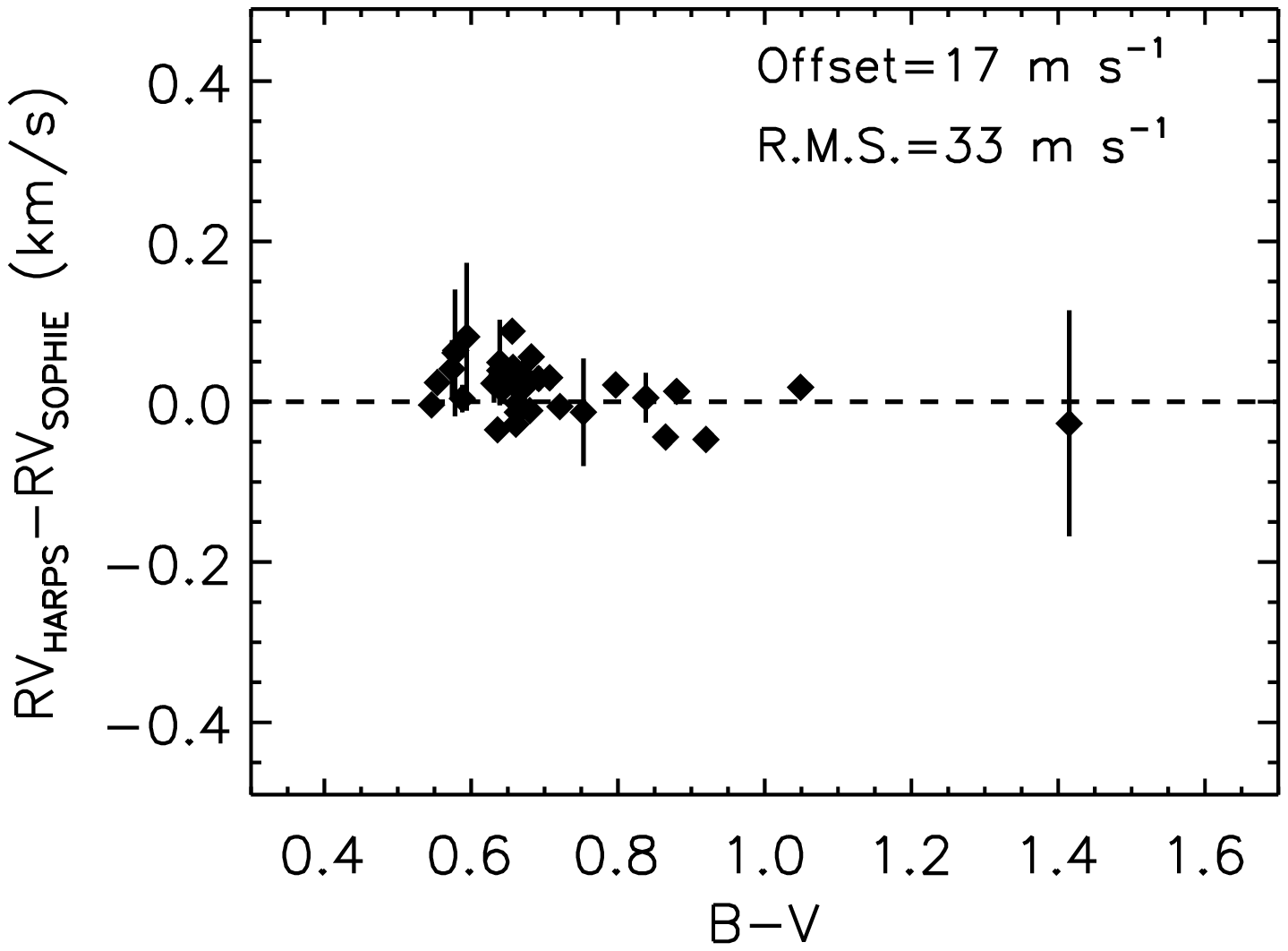}
\caption{Offsets between RV measurements from \sophie\ and the other instruments as provided in Table~\ref{t:offsets}. Error bars when available,  are the quadratic sum of the two standard deviations obtained for a given star observed several times with the two instruments being considered. }
\label{f:offsets}
\end{center}
\end{figure*}

Figure~\ref{f:offsets} shows the RV offsets between the various instruments versus the $B-V$ colour index. The error bars, when available, 
are the quadratic sum of the two standard deviations obtained for a given star observed several times with the two considered instruments. A large error bar is an indication that the star is probably variable.
The offsets are constant in general, except for \elodie\ where a dependency on colour is clearly seen. 
The scatter is also much higher when \elodie\ is involved. The scatter reflects the RV variations, expected to be more pronounced in the case of a secular drift as the time baseline of the measurements increases, which is the case  when \elodie\ observations are considered. The scatter also reflects 
the RV errors, which are higher for \elodie\ than for the other instruments (see Table~\ref{t:nb_mes}). Table~\ref{t:offsets} summarizes the mean difference and RMS scatter 
obtained when comparing the measurements from two instruments.  In this process, an iterative clipping at the $3\sigma$
level is performed to remove the outliers.  The offset column provides the velocity 
 corrections applied to the measurements of \coralie, \narval, \elodie, and \harps\ to translate them onto the \sophie\ scale. 
 
\begin{figure}[h!]
\begin{center}
 \includegraphics[width=0.5\textwidth]{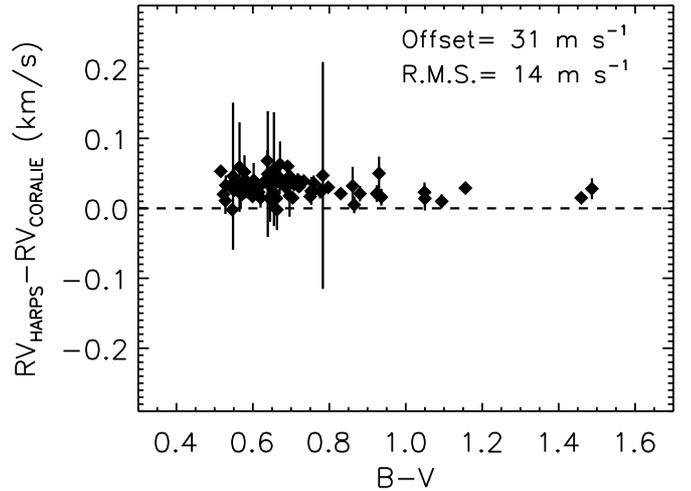}        
\caption{Offsets between RV measurements from \harps\ and \coralie\ showing a remarkably low RMS scatter.}
\label{f:off_coha}
\end{center}
\end{figure}

 \citet{boi12} have also measured the RV offset between \sophie\ and \elodie\ and found a steeper slope ($-425.6(B-V)+202.4$) but their fit is limited to $B-V < 0.75$. Table~\ref{t:offsets} and Fig.~\ref{f:off_coha} also present the comparison of the \coralie\ and \harps\ measurements as an illustration of the excellent agreement between those two instruments.
 
\begin{table}[h]
  \centering 
  \caption{RV offsets between instruments derived for stars in common. }
  \label{t:offsets}
\begin{tabular}{|l|r|r|c|}
\hline
Instruments & $N_{\rm common}$      & $N_*$	 & Offset  \\
                         &     &         (a)         &  (\ms)  \\
\hline
  \coralie$-$\sophie & 164 & 154  & -23  $\pm$ 2.4 \\
 \narval$-$\sophie & 114  & 108  & -26  $\pm$ 6.2 \\
 \elodie$-$\sophie  &277  & 247  &  (b)  \\
 \harps$-$\sophie &  36   & 34   & 17  $\pm$  5.0 \\
 \harps$-$\coralie& 88  &  84 &  31 $\pm$ 1.5 \\
 \hline
\end{tabular}
\tablefoot{(a) $N_*$ is the number of stars effectively used after  an iterative clipping at a 3$\sigma$ level. \\
(b) In the case of \elodie$-$\sophie\ a linear regression was performed, giving an offset as a function of the colour index Offset$=-259.0(B-V)+105.2 \pm 2.9$ \ms.}
\end{table}

Two tables are provided for the catalogue, only available electronically. Table \ref{t:indiv_meas}  gives 
the original individual RV measurements, and Table~\ref{t:catalog} gives for each star :

\begin{itemize}
\item the basic data from Hipparcos: HIP number, equatorial coordinates, V magnitude, $B-V$, spectral type;
\item the catalogue(s) from which the star was selected : \citet{nid02}, \citet{fam05}, \citet{nor04} or IAU standard \citep{udr99};
\item the weighted mean radial velocity in the \sophie\ scale, $\rm \overline{RV_S}$, with a weight $w_i$ applied to
 each individual velocity measurement RV$_i$ being $w_i=1/\epsilon_i'^2$ and $\epsilon_i'$ the quadratic sum of $\epsilon_i$ and the offset error listed in Table~\ref{t:offsets};
 \item the internal error of $\rm \overline{RV_S}$ : $I=\underset{i}{\sum} w_i\, \epsilon_i' /\underset{i}{\sum} w_i$;
\item the  $\rm \overline{RV_S}$ weighted standard deviation $\sigma_{\rm RV_S}$ defined as 
$$\sigma_{\rm RV_S}^2=\frac{\underset{i}{\sum} w_i}{(\underset{i}{\sum} w_i)^2 - \underset{i}{\sum} w_i^2 }\,
\underset{i}{\sum} w_i ({\rm RV}_i -{\rm \overline{RV_S}})^2$$
\item the number of observations $N$;
\item the $\rm \overline{RV_S}$ uncertainty defined as the maximum of the standard error $\sigma_{\rm RV_S} / \sqrt N$ and $I / \sqrt N$ \citep{jas88};
\item the time baseline in days;
\item the mean Julian day of the $N$ observations.
\end{itemize}

\begin{table}[h]
  \centering 
  \caption{Excerpt of the catalogue table giving the 10214 original individual RV measurements and 
their error $\epsilon$ (see Sect.~\ref{sec:measurements}), the Julian day of the observation, the instrument, and mask used. The name of the instrument is abbreviated with its first letter.}
  \label{t:indiv_meas}
\begin{tabular}{cccccc}
\hline
ID &  Instr.  & JD    & RV	 & $\epsilon$ & mask \\
    &    & -2400000    & \kms	 & \kms & \\
 \hline
HIP000296   &   C &52173   &10.932  &0.0070 &G2 \\
 HIP000296   &   C &53262   &10.930  &0.0072 &G2 \\
 HIP000296    &  C &54354   &10.952  &0.0064 &G2 \\
 HIP000296    &  C &55167   &10.948  &0.0109 &G2 \\
 HIP000407   &   C &51542   &12.346  &0.0093 &G2 \\
 HIP000407    &  C &52649   &12.316  &0.0069 &G2 \\
 HIP000407   &   C &53921   &12.328  &0.0075 &G2 \\
 HIP000407   &   C &55167   &12.331  &0.0123 &G2 \\
 HIP000420   &   C &51382   &34.861  &0.0082 &G2 \\
 HIP000420   &   C &53218   &34.852  &0.0083 &G2 \\
 HIP000420   &   C &55167   &34.868  &0.0130 &G2 \\
 HIP000420    &  C &55758   &34.879  &0.0068 &G2 \\
 HIP000466     & S &55131   &-8.371  &0.0080 &K5 \\
 HIP000466    &  S &55452   &-8.364  &0.0080 &K5 \\
 HIP000556    &  C &55882    &0.109  &0.0110 &G2 \\
 HIP000556    &  C &56221    &0.091  &0.0135 &G2 \\
 HIP000556    &  S &55129    &0.102  &0.0086 &G2 \\
\hline
\end{tabular}
\end{table}

\begin{table*}[h]
\tiny
  \centering 
  \caption{Excerpt of the catalogue table giving the mean RV on the \sophie\ scale of  the 1420 standard star candidates  (see text for the description of the columns). Abbreviations are used for the catalogue(s) from which the star was selected: ni=\citet{nid02}, fa=\citet{fam05}, nor=\citet{nor04} in column 6.}
  \label{t:catalog}
\begin{tabular}{cccclrrccrcrc}
\hline
ID &  $\alpha, \delta$ & Vmag &  $B-V$ & ST & cat &$\rm \overline{RV_S}$  &  I &  $\sigma_{\rm RV_S}$ &  N & uncertainty  & span & mean JD \\
    &   & & &  &   &  \kms &     \kms&    \kms& & \kms &  days & - 2400000\\
\hline
 HIP000296  & 00:03:41.30 -28:23:45.1 & 8.24 & 0.780 &G8V       &     nor &  10.963 &  0.0078  & 0.0117  &   4 &  0.0059 & 2994 &53739 \\
 HIP000407  & 00:04:58.63 -70:12:43.9 & 8.13 & 0.710 &G5V       &     nor &  12.351 &  0.0086  & 0.0130  &   4 &  0.0065 & 3625 &53319 \\
 HIP000420  & 00:05:07.60 -52:09:05.1 & 7.53 & 0.577 &G0V       &     nor &  34.889 &  0.0086  & 0.0129  &   4 &  0.0065 & 4376 &53881 \\
 HIP000466  & 00:05:34.91 +53:10:18.1 & 7.23 & 1.174 &K0        & fa      &  -8.367 &  0.0080  & 0.0049  &   2 &  0.0057 &  321 &55291 \\
 HIP000556  & 00:06:46.94 -04:20:59.3 & 8.21 & 0.574 &F8        &     nor &   0.113 &  0.0104  & 0.0165  &   3 &  0.0095 & 1092 &55744 \\
 HIP000616  & 00:07:32.34 -23:49:08.2 & 8.70 & 0.798 &K0V       &   ni    & -42.994 &  0.0086  & 0.0049  &   4 &  0.0043 & 3380 &53629 \\
 HIP000624  & 00:07:37.41 -45:07:09.5 & 8.19 & 0.584 &F8/G0V    &     nor &  14.802 &  0.0117  & 0.0134  &   2 &  0.0095 &  725 &55145 \\
 HIP000699  & 00:08:41.02 +36:37:38.7 & 6.21 & 0.504 &F8IV      &   ninor & -15.116 &  0.0043  & 0.0119  &  16 &  0.0030 & 5160 &54912 \\
\hline
\end{tabular}
\end{table*}

\begin{figure}[ht!]
\begin{center}
 \includegraphics[width=\columnwidth]{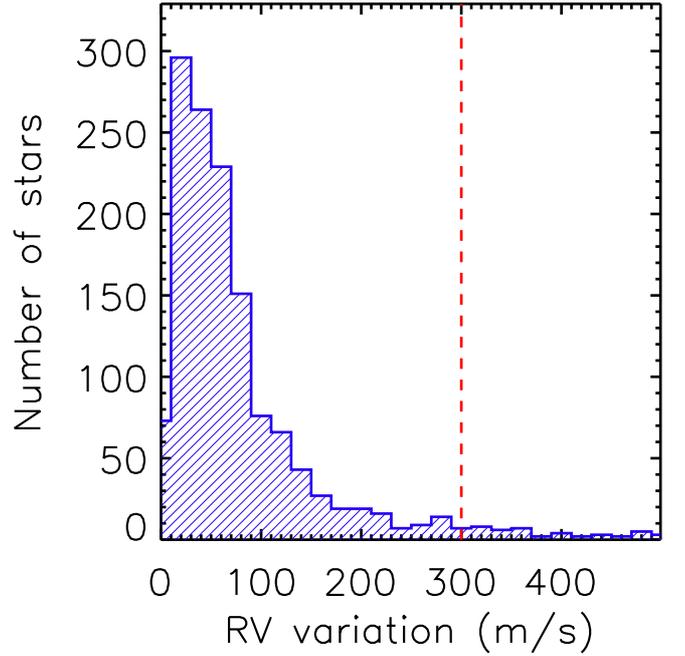}      
 \caption{Distribution of RV variations of the candidate standard stars having at least two RV measurements separated 
 by  100 days or more. The variation is defined by $3\sigma_{\rm RV_S}$. A dashed line shows the 300 \ms\ stability threshold adopted for the calibration of the RVS instrument.}.
\label{f:histo_var}
\end{center}
\end{figure}

Figure~\ref{f:histo_var} shows the histogram of the variations of the 1420 candidates. We define the level of stability of a given star by $3\sigma_{\rm RV_S}$. We find that 92.8\% of the stars have a stability better than 300 
m s$^{-1}$,  which is the threshold defined for the calibration of the RVS instrument. More than 1000 stars exhibit a stability better than 100 \ms\ (i.e. $\sigma_{\rm RV_S} < 33$ \ms). 

\section{Comparison with other catalogues}
\label{sec:comparison}
\begin{figure*}[ht!]
\begin{center}
 \includegraphics[width=0.5\textwidth]{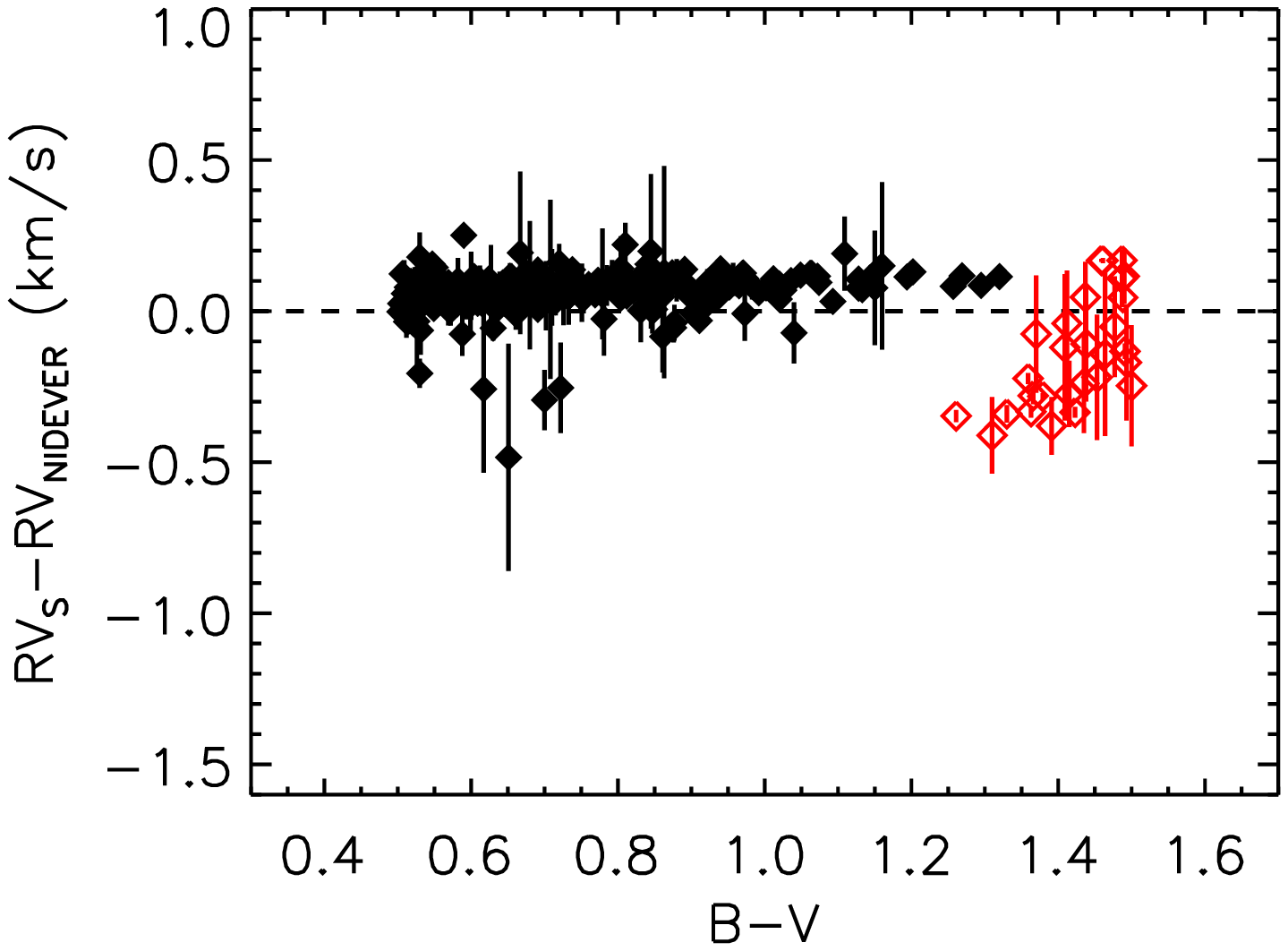}\includegraphics[width=0.5\textwidth]{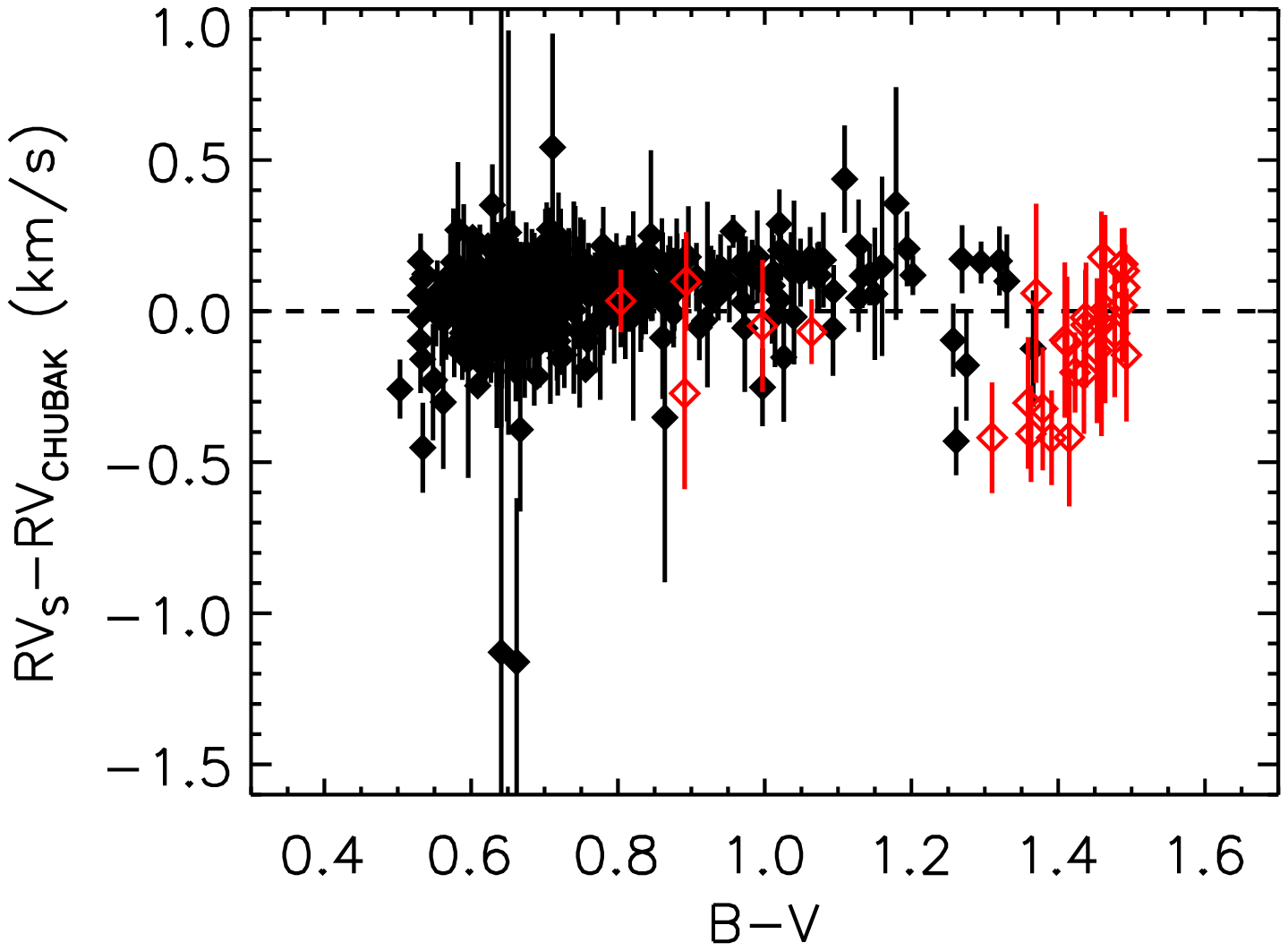}     
 \caption{Difference between our radial velocities and those from \citet{nid02}  (left panel) and  
 \citet{chu12}  (right panel) as a function of the $B-V$ colour index. Red dots indicate the stars they have analysed with an M template, while for the other ones they used the solar spectrum as reference.  The offset and RMS values are given for FGK stars and for M stars in Table~\ref{t:chub-nid}.
 }
\label{f:nid-chub}
\end{center}
\end{figure*}

The catalogue of \citet{nid02} contains  889 late-type stars followed during several years, with 782 of them exhibiting a velocity scatter less than 100 \ms.
Some 336 stars of the catalogue presented here have been selected from the stable stars of \citet{nid02} and fulfil the Gaia-RVS requirements.  Our catalogue also has 355 stars in common with the recent list of \citet{chu12} which provides RVs for 2046 nearby FGKM stars. Comparisons with those two catalogues are shown in Fig.~\ref{f:nid-chub} as a function of $B-V$.
 There is general good agreement with small zero-point differences. The outliers reveal the intrinsic variable stars, as well as systematics due to methodology. 
   For $B-V \gtrsim 1.2$  there is a clear systematic effect. We note that 
    both \citet{nid02} and \citet{chu12} use the same technique of chi-square minimization to determine the relative shifts between a template spectrum and an observed spectrum, but not in the same spectral range. They use an M-dwarf observed spectrum as template for cool stars and a solar spectrum for the other stars. The two catalogues are constructed on the same scale, as explained in \citet{chu12}.
We divide the comparison of their values to the present ones into two parts, according to their reference template. Results are presented in Table~\ref{t:chub-nid}.

\begin{table}[h]
  \centering 
  \caption{Comparison of RVs obtained with by \citet{nid02} and \citet{chu12} and those presented here, according to the templates they used. 
  Differences, $\Delta_{\rm RV} = {\rm RV_S}-\rm RV_{\rm CHUBAK / NIDEVER}$ (Fig.~\ref{f:nid-chub}) and RMS scatter in \ms\  are derived with an iterative clipping at a 3$\sigma$ level. N$_*$ is the number of stars in common, N$_o$ is the number of outliers removed.}
  \label{t:chub-nid}
\begin{tabular}{|l|l|r|r|}
\hline
source & stat & Sun template      & M-dwarf template \\
\hline
 \multirow{4}{*}{Chubak et al.}& N$_*$ & 326 & 29 \\
& N$_o$ & 15 & 0 \\
 & $\Delta_{\rm RV}$ & 63  &  -98 \\
 & RMS & 100 & 182 \\
 \hline
 \multirow{4}{*}{Nidever et al.} & N$_*$ & 308 & 28 \\
 & N$_o$ & 22 & 0 \\
& $\Delta_{\rm RV}$ & 72 &  -141 \\
& RMS & 33 & 178 \\
  \hline
\end{tabular}
\end{table}

We get very similar zero-point difference in both cases, which is expected since the catalogues of \citet{nid02} and \citet{chu12} are on the same scale. 
The very low RMS scatter of 33  \ms\ obtained for the comparison to  \citet{nid02} for FGK stars  is remarkable  for two reasons. First, since the median epoch of Nidever et al. observations (1999.0) is earlier than ours (2006.9), it means that most of the stars in common have not varied much during that period. Second, it also shows that the precision of the present catalogue is not worse than 33 \ms. The same level of scatter has already been measured by \citet{nid02} when they compared their catalogue to the best set of standard stars available at the time, assembled by the Geneva team \citep{udr99} and based on {\tiny CORAVEL} and \elodie. The \elodie\ measurements that were used to build the Udry et al. standard star catalogue are among those retrieved from the archive for the present catalogue. We do not show the comparison to \citet{udr99} since we have measurements in common. It is worth noticing that there is no correlation of the differences with \citet{nid02} with $B-V$ in the  FGK range, even though the stars of the present catalogue have been analysed with different masks.

For RVs obtained  by  \citet{chu12} and \citet{nid02} with an M-dwarf template, the mean difference and RMS are much larger. Figure~\ref{f:nid-chub} suggests that there might be a correlation of these differences with colour, but the $B-V$ interval is too narrow to confirm this claim. Such significant offset and RMS compared with  \citet{udr99} have already been found and discussed in both articles, so we confirm with a larger sample the assertion by \citet{chu12} that {\it 'the M dwarf velocities in general, from all surveys, remain uncertain at the level of 200 \ms\ (RMS) and
harbor uncertain zero points at the level of 150 \ms'}.

\begin{figure}[ht!]
\begin{center}
 \includegraphics[width=\columnwidth]{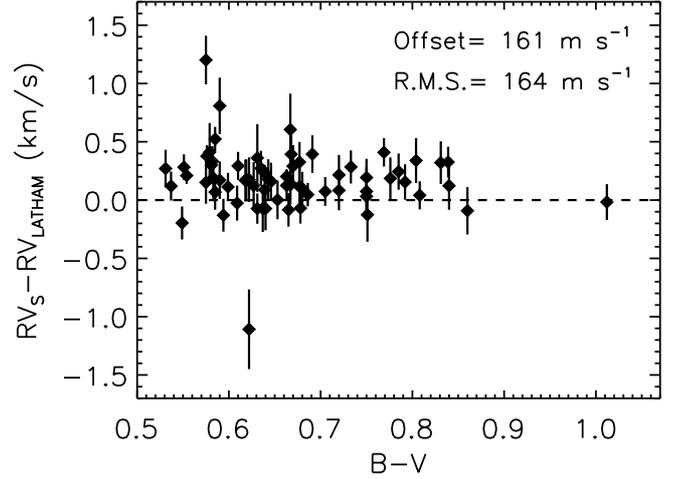}      
 \caption{Comparison of RVs obtained with by \citet{lat02}  and those presented here. The offset and RMS values have been obtained after rejectiing two outliers.}.
\label{f:latham}
\end{center}
\end{figure}

The comparison with the catalogue of \citet{lat02} is shown in Fig.~\ref{f:latham}.  This catalogue contains 1359 single-lined stars on the 
Center for Astrophysics (CfA) scale. The difference is: $\Delta_{\rm RV} = {\rm RV_S}-\rm RV_{\rm LATHAM}=161$ \ms\ for 66 stars in common after rejection of two outliers. This offset is quite significant but it agrees with the zero-point correction of 139 \ms\ that they determined from asteroids. Although 
the Latham et al. RVs are presented as having errors ranging between 0.5 to 
1.5 \kms, we measure a much lower RMS scatter, 164 \ms, which suggests that 
 their errors are lower for the 65 common stars.

\section{Stable and variable stars}
\label{s:stable_stars}
\begin{figure}[ht!]
\begin{center}
 \includegraphics[width=\columnwidth]{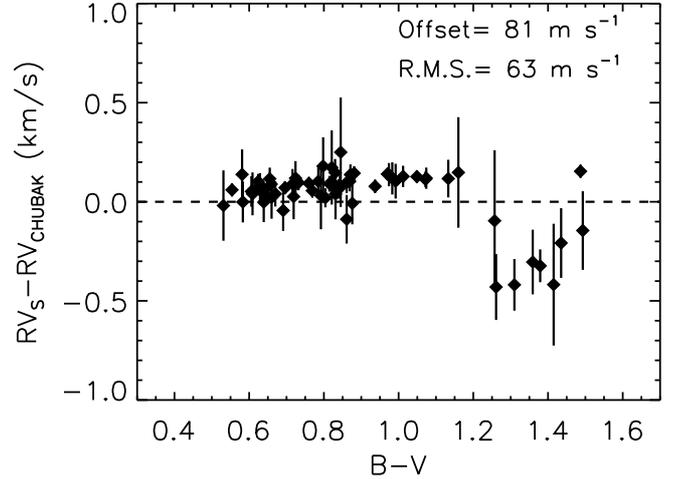}      
 \caption{Difference in radial velocities of the Chubak et al. standard stars for 63 stars in common. The offset and RMS values have been obtained after rejecting seven outliers at $B-V >$ 1.2.}
\label{f:std_chub}
\end{center}
\end{figure}

\citet{chu12} present a list of 131 FGKM standard stars having constant radial velocity over ten years.  We have 63 stars in common with this list and the comparison is presented in Fig.~\ref{f:std_chub}. The average offset is 
$\Delta_{\rm RV} = {\rm RV_S-RV_{Chubak}} = 81$ \ms, with an RMS scatter of 63 \ms\ after rejection of seven outliers at $B-V \>$ 1.2. This RMS scatter is lower than in the comparison with the main catalogue of \citet{chu12}  presented in Table~\ref{t:chub-nid}, but does not reach the very low value of 33 \ms\ obtained in the comparison with \citet{nid02}. This means that either the Chubak et al. standard stars are less stable than the Nidever et al. ones, or that  their errors are significantly higher than the Nidever et al. ones.

The catalogue presented here is also a good source for very stable stars that are good candidates for becoming official RV-standards useful for many projects other than Gaia. Table~\ref{t:catalog} can easily be queried with  criteria such as 

\begin{itemize}
\item  at least 4 measurements available,
\item standard deviation on  the \sophie\ scale $\sigma_{\rm RV_S} < 33$ \ms\ (i.e. stability better than 100 \ms),
\item time baseline of ten years at least.
\end{itemize}

\begin{figure*}[h!]
\begin{center}
\includegraphics[width=0.5\textwidth]{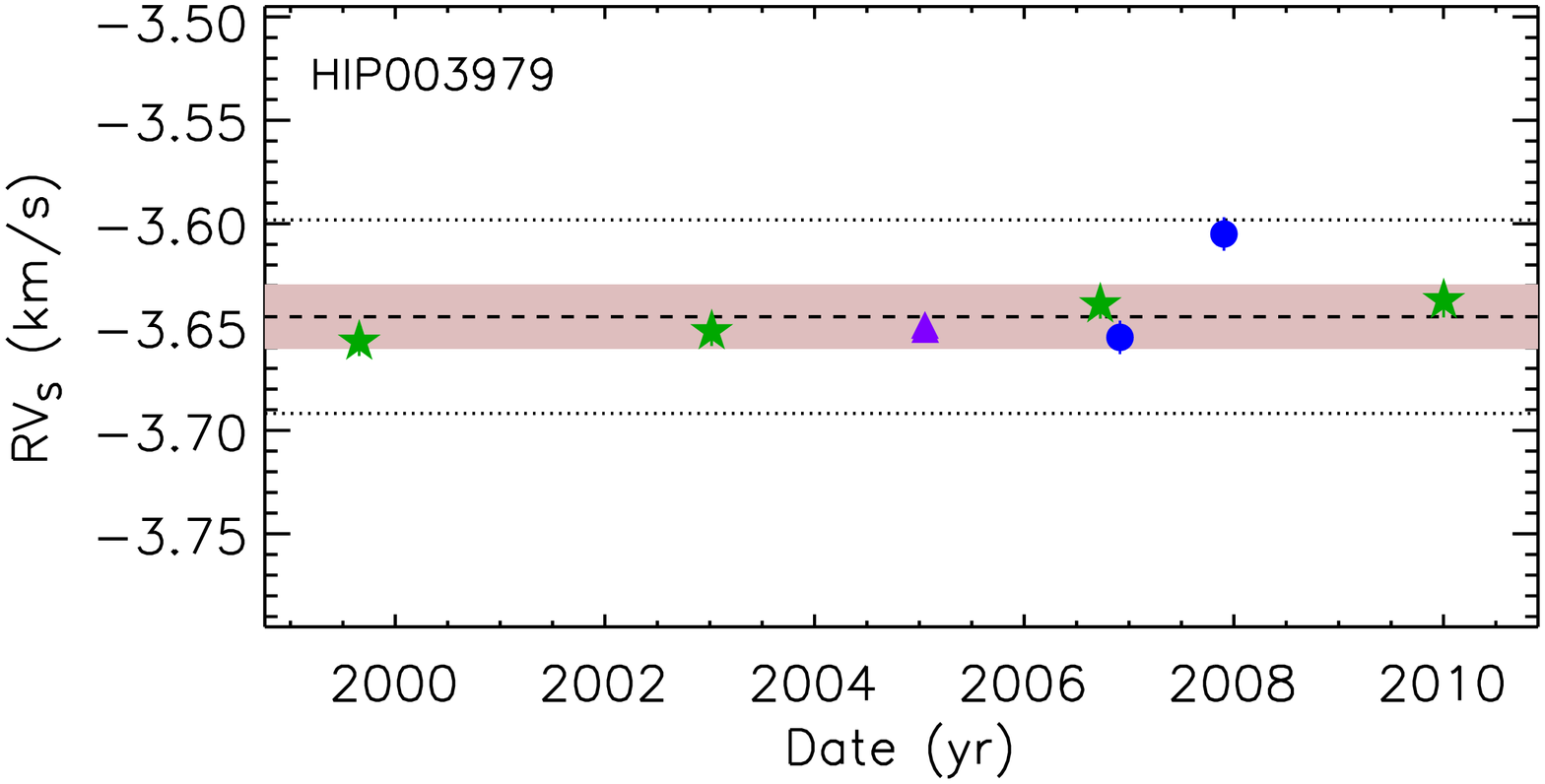}\includegraphics[width=0.5\textwidth]{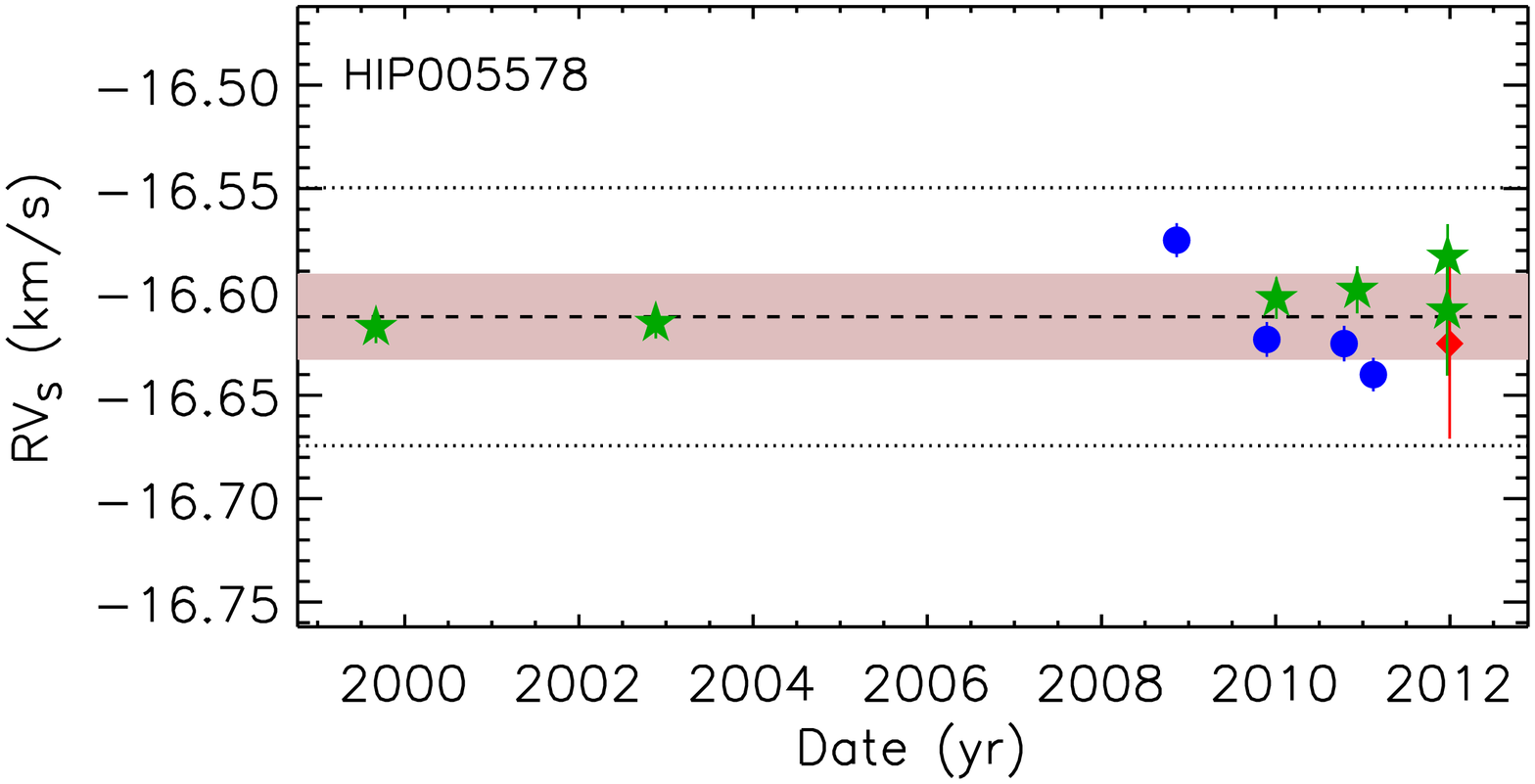}      
 \includegraphics[width=0.5\textwidth]{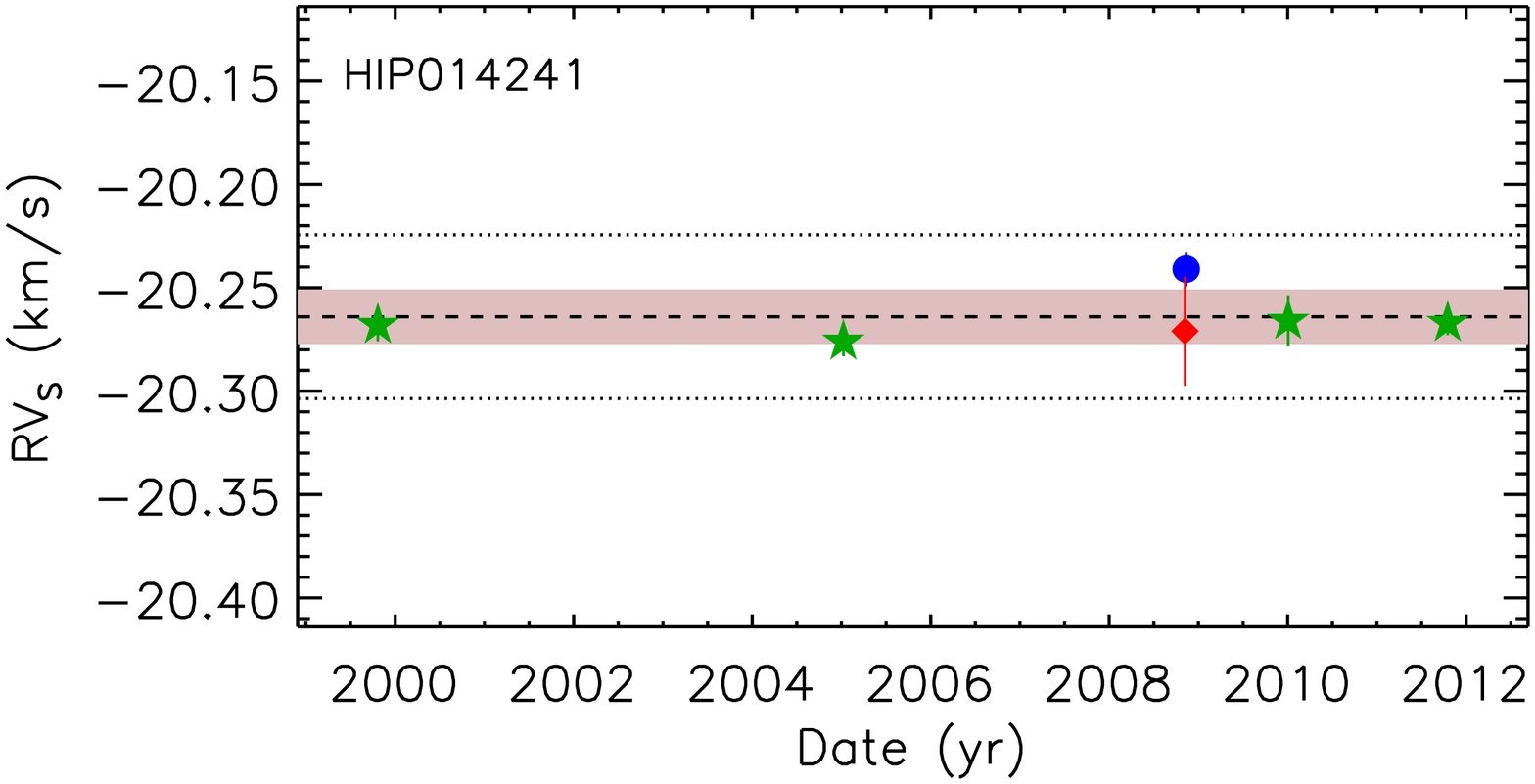}\includegraphics[width=0.5\textwidth]{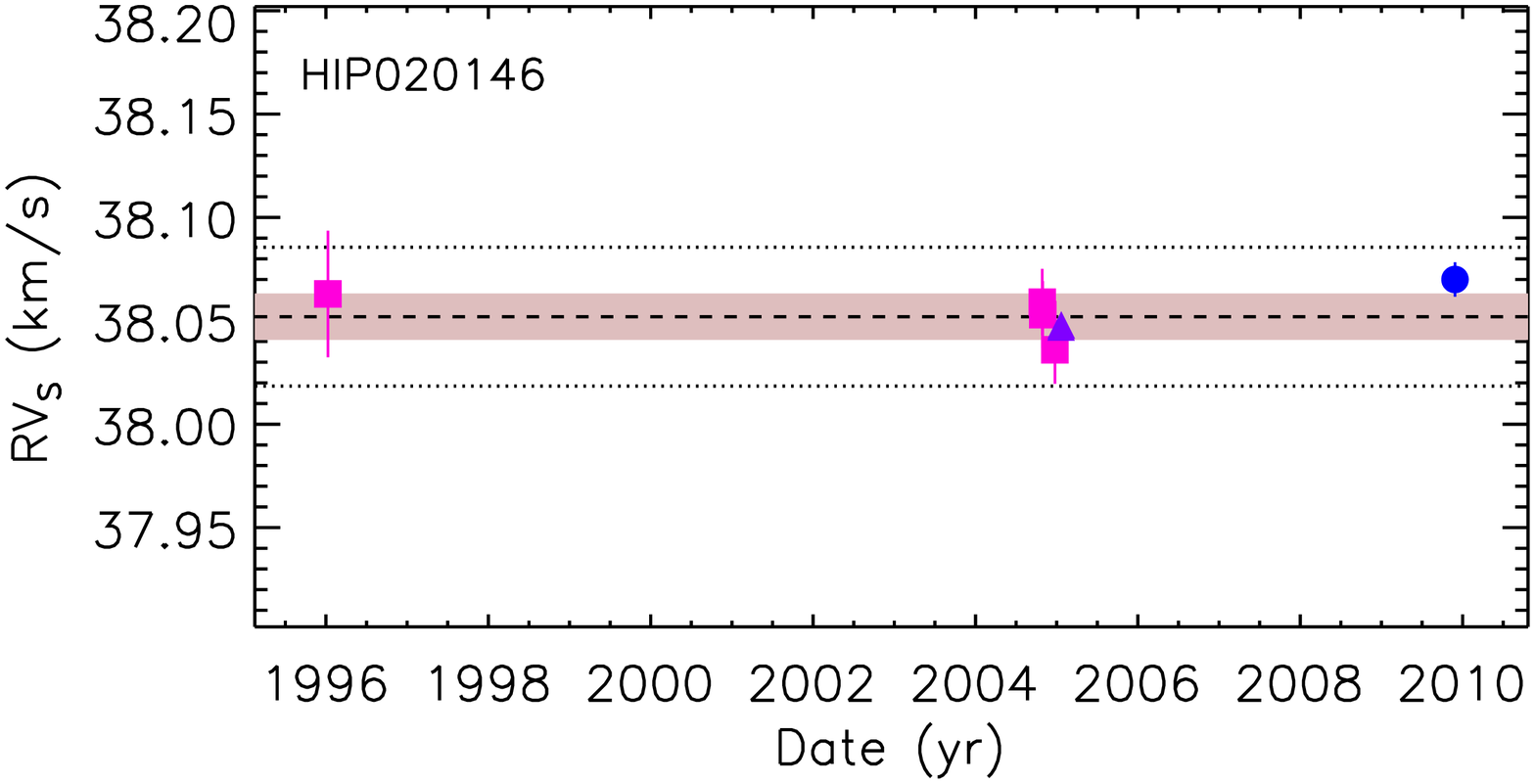}
\includegraphics[width=0.5\textwidth]{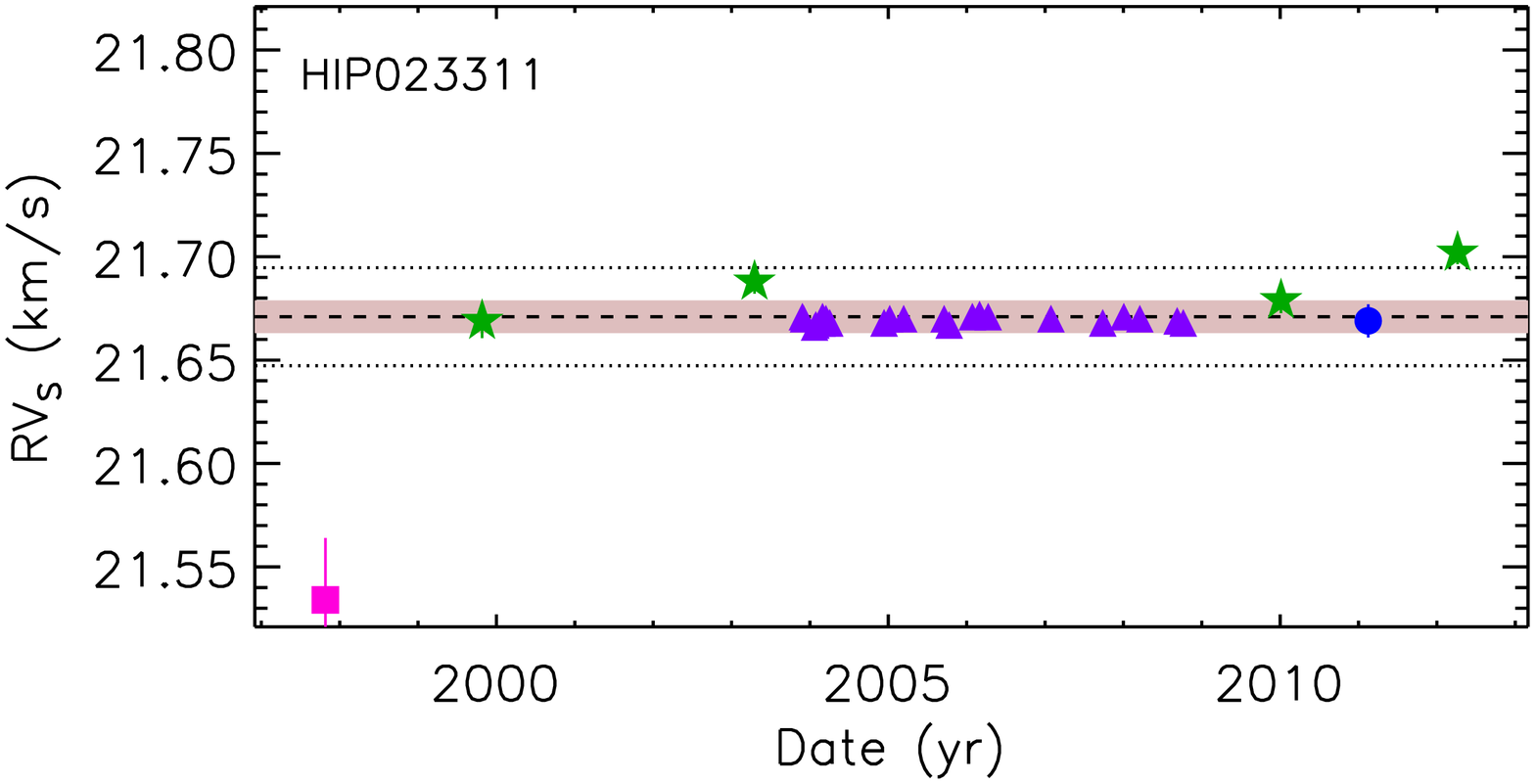}\includegraphics[width=0.5\textwidth]{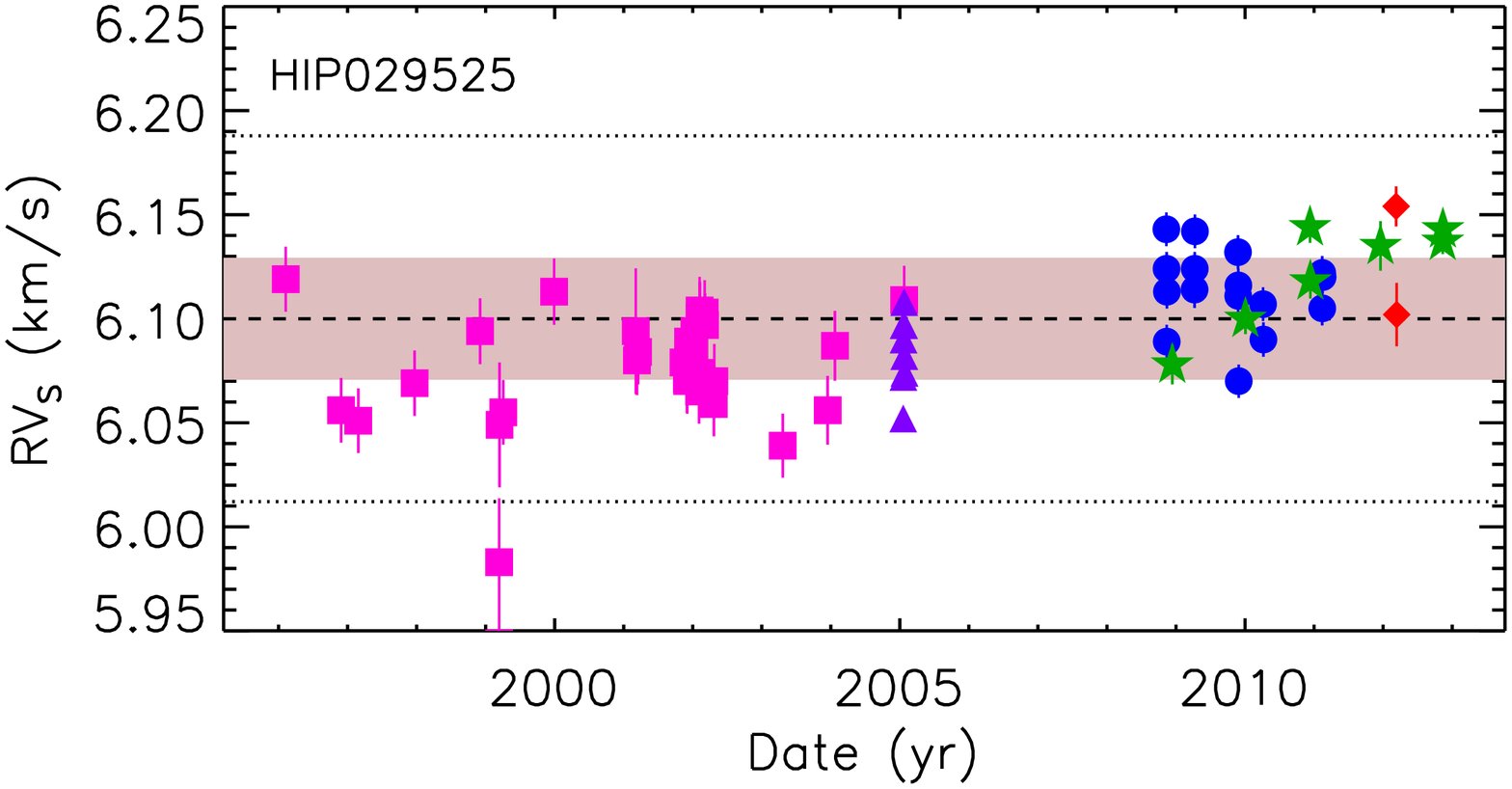}
\includegraphics[width=0.5\textwidth]{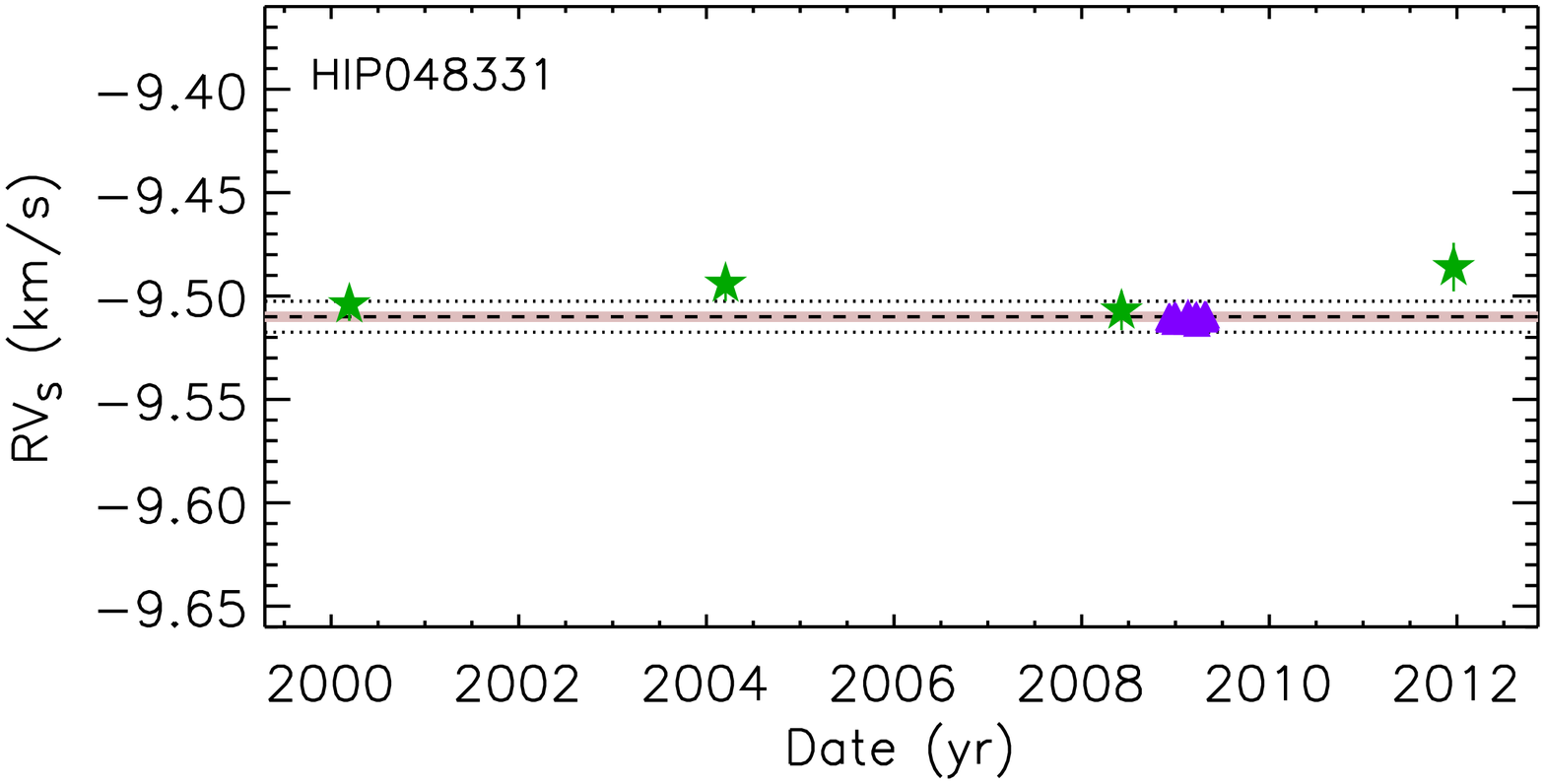}\includegraphics[width=0.5\textwidth]{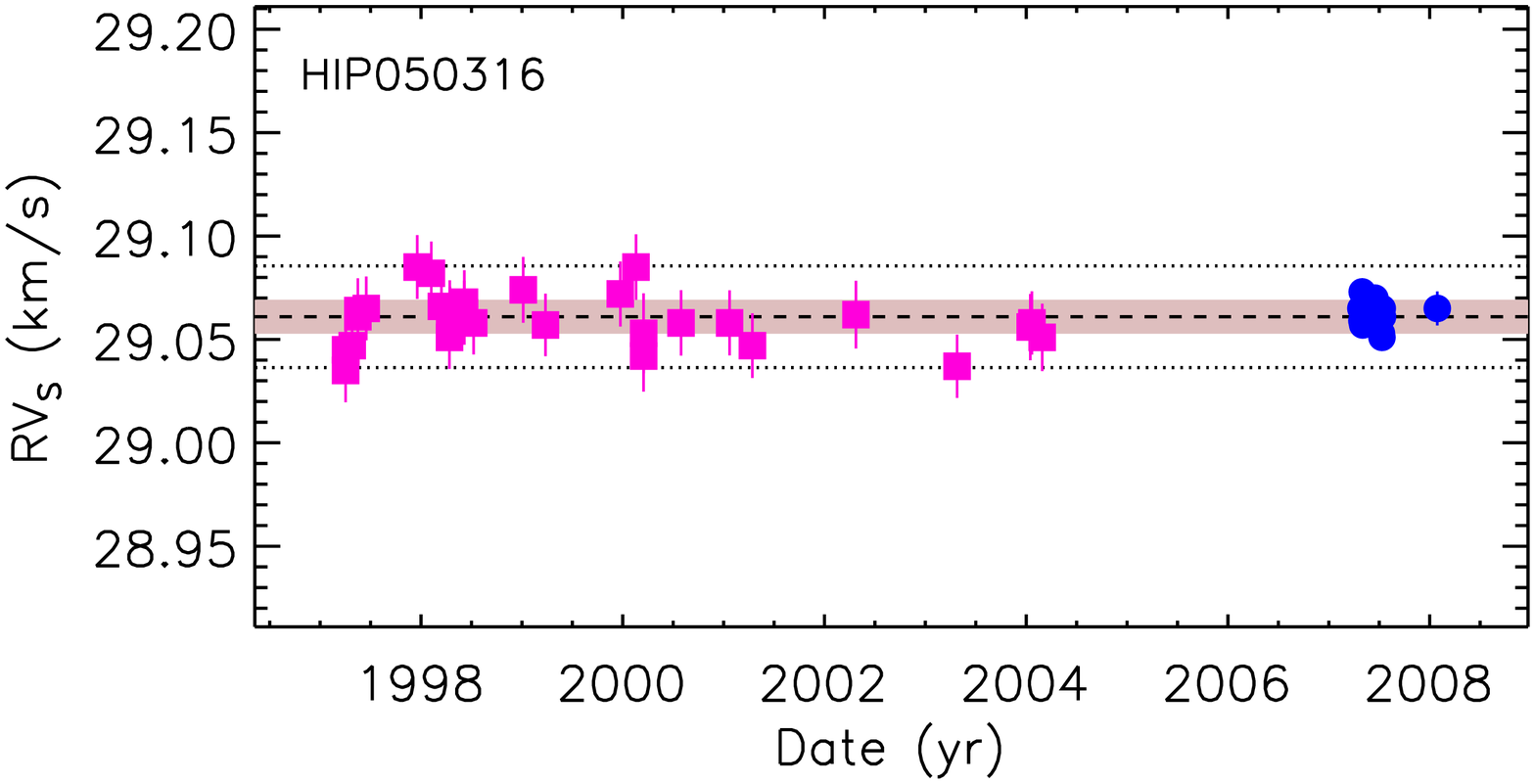}
\includegraphics[width=0.5\textwidth]{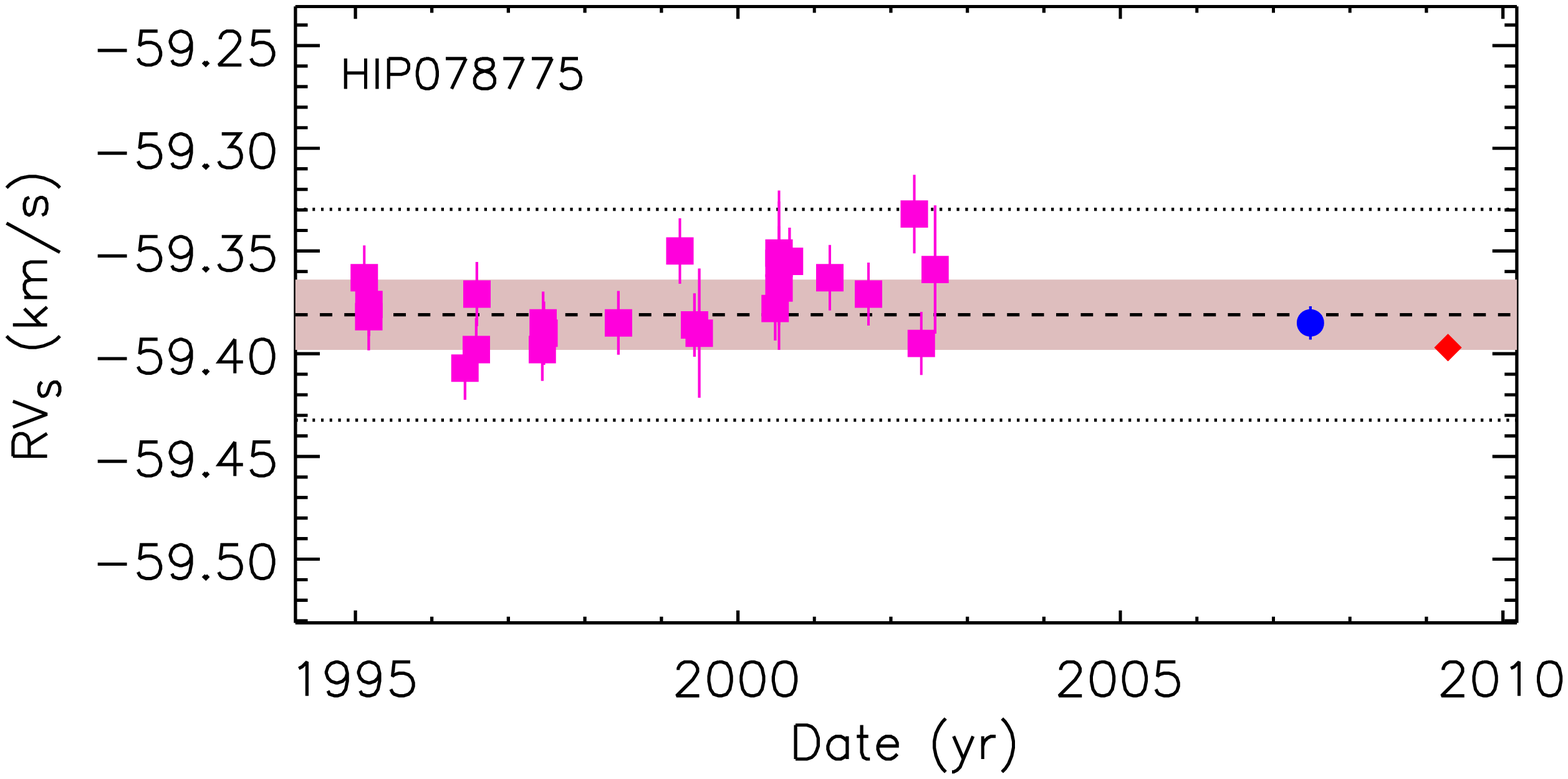}\includegraphics[width=0.5\textwidth]{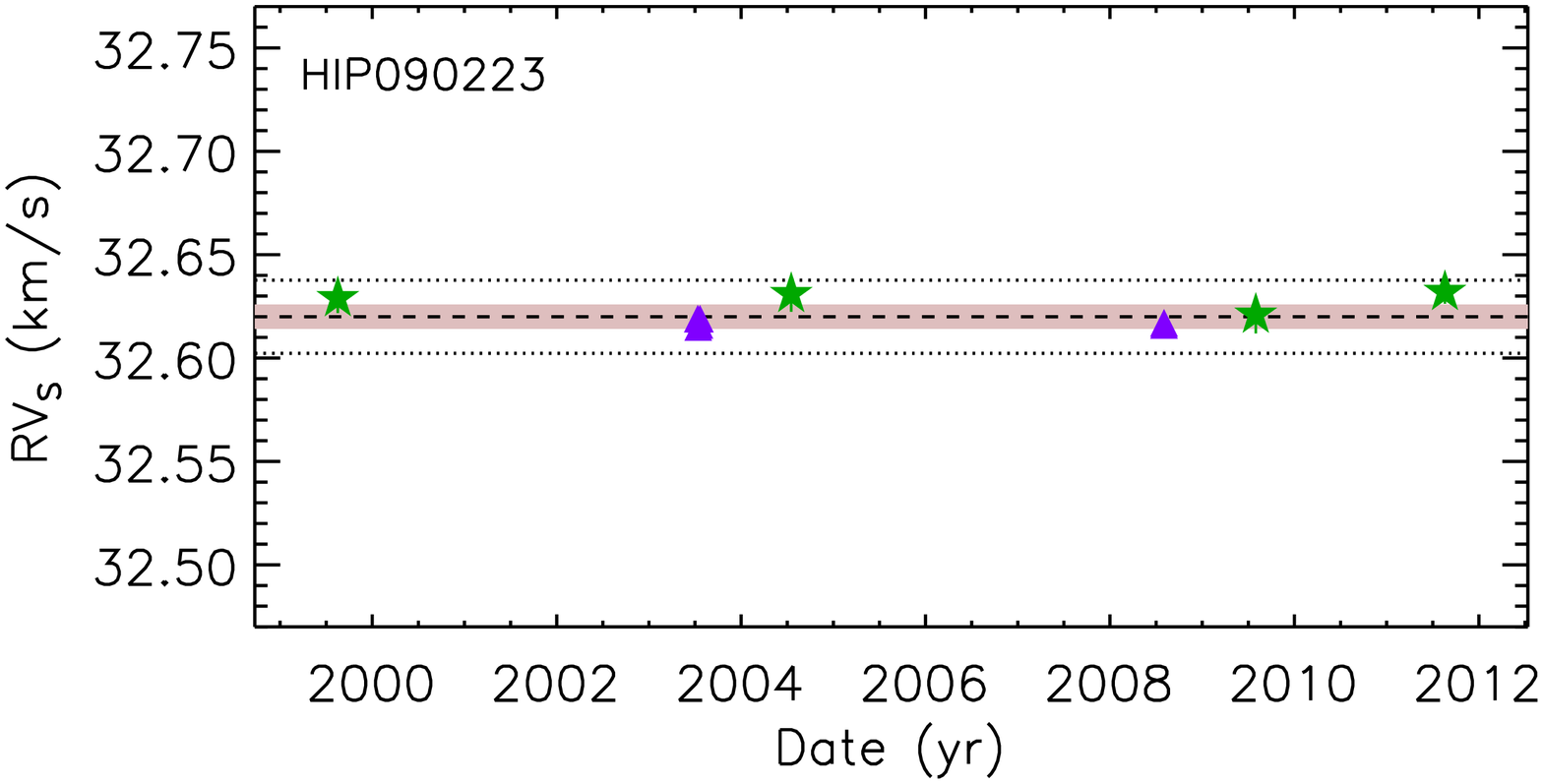}
\caption{RV measurements for some of the long-term stable stars of the catalogue. The RV axis is centred on $\overline {\rm RV_S}$ and spans 300 \ms. The shaded area represents 1-$\sigma_{\rm RV_S}$ and the dotted lines 3-$\sigma_{\rm RV_S}$. Blue dots represent \sophie\ mesurements, green stars \coralie, pink squares \elodie, purple triangles \harps, red diamonds \narval.}
\label{f:stable_stars}
\end{center}
\end{figure*}

More than 300 stars verify these conditions and should be followed up in the future to become official RV-standards. Several of these stars are shown in Fig.~\ref{f:stable_stars}. They were observed with  different instruments  and the agreement of the measurements demonstrates that the zero points  have been properly corrected. 

\begin{figure*}[h!]
\begin{center}
 \includegraphics[width=0.5\textwidth]{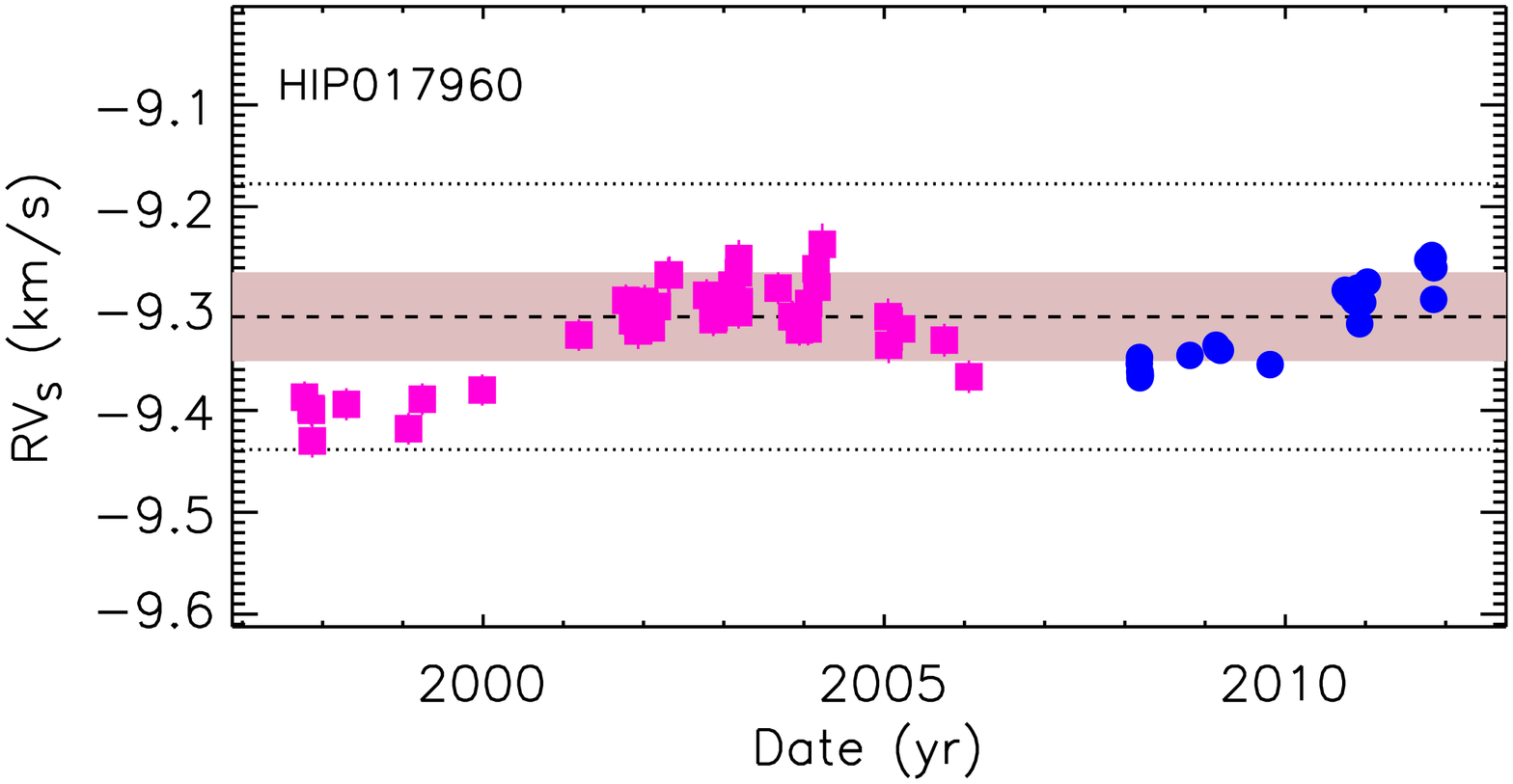}\includegraphics[width=0.5\textwidth]{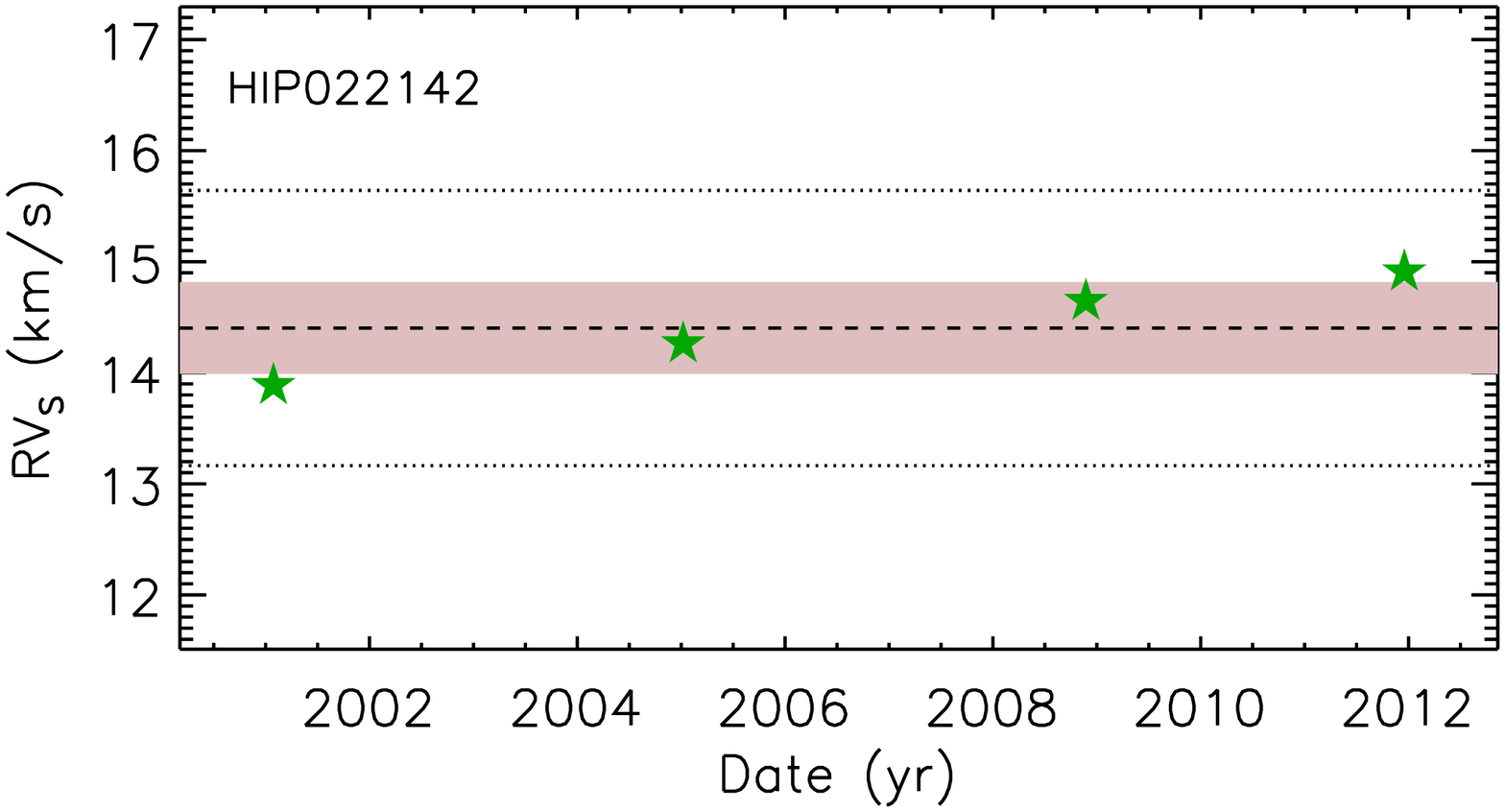}
\includegraphics[width=0.5\textwidth]{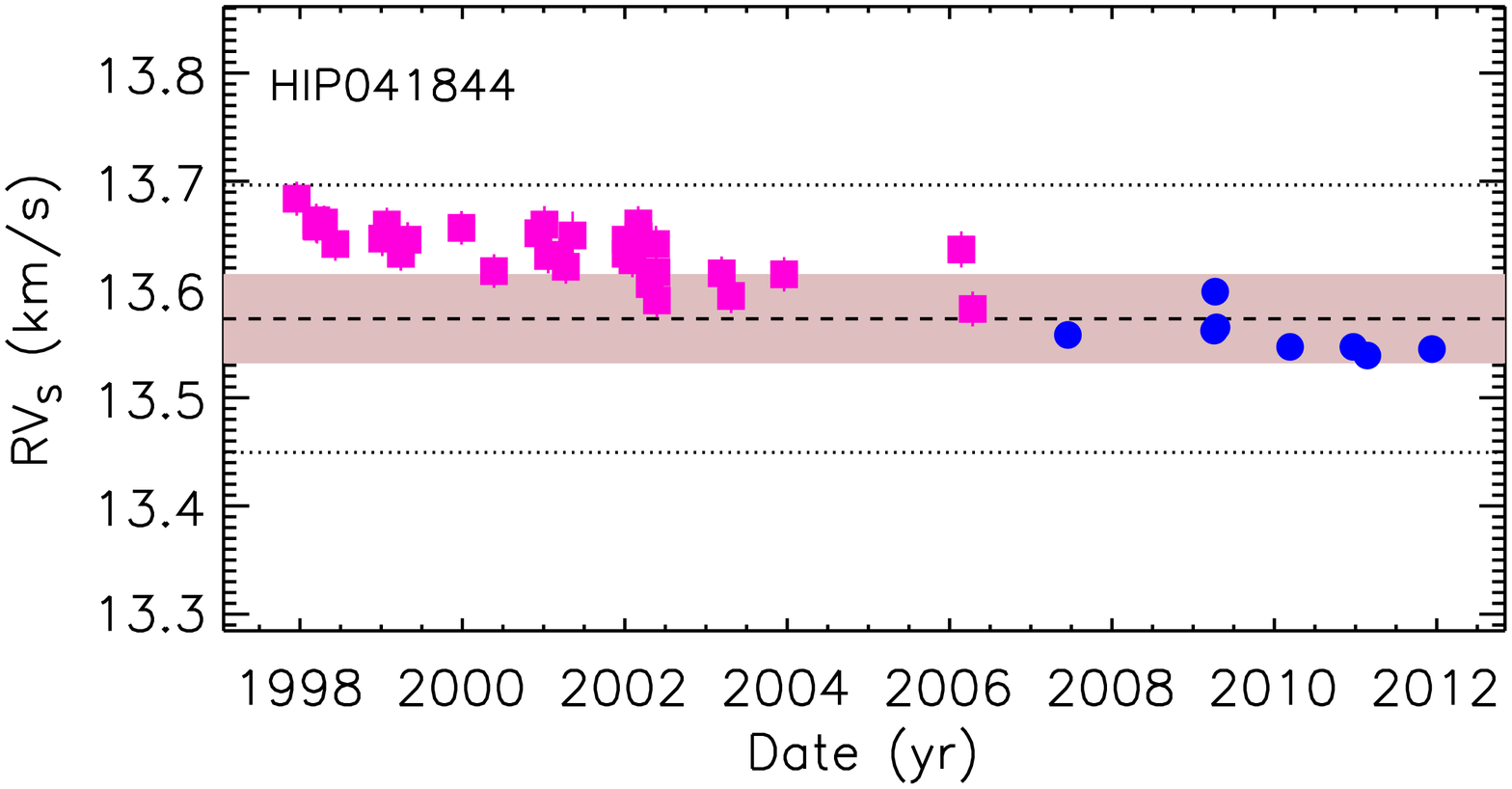}\includegraphics[width=0.5\textwidth]{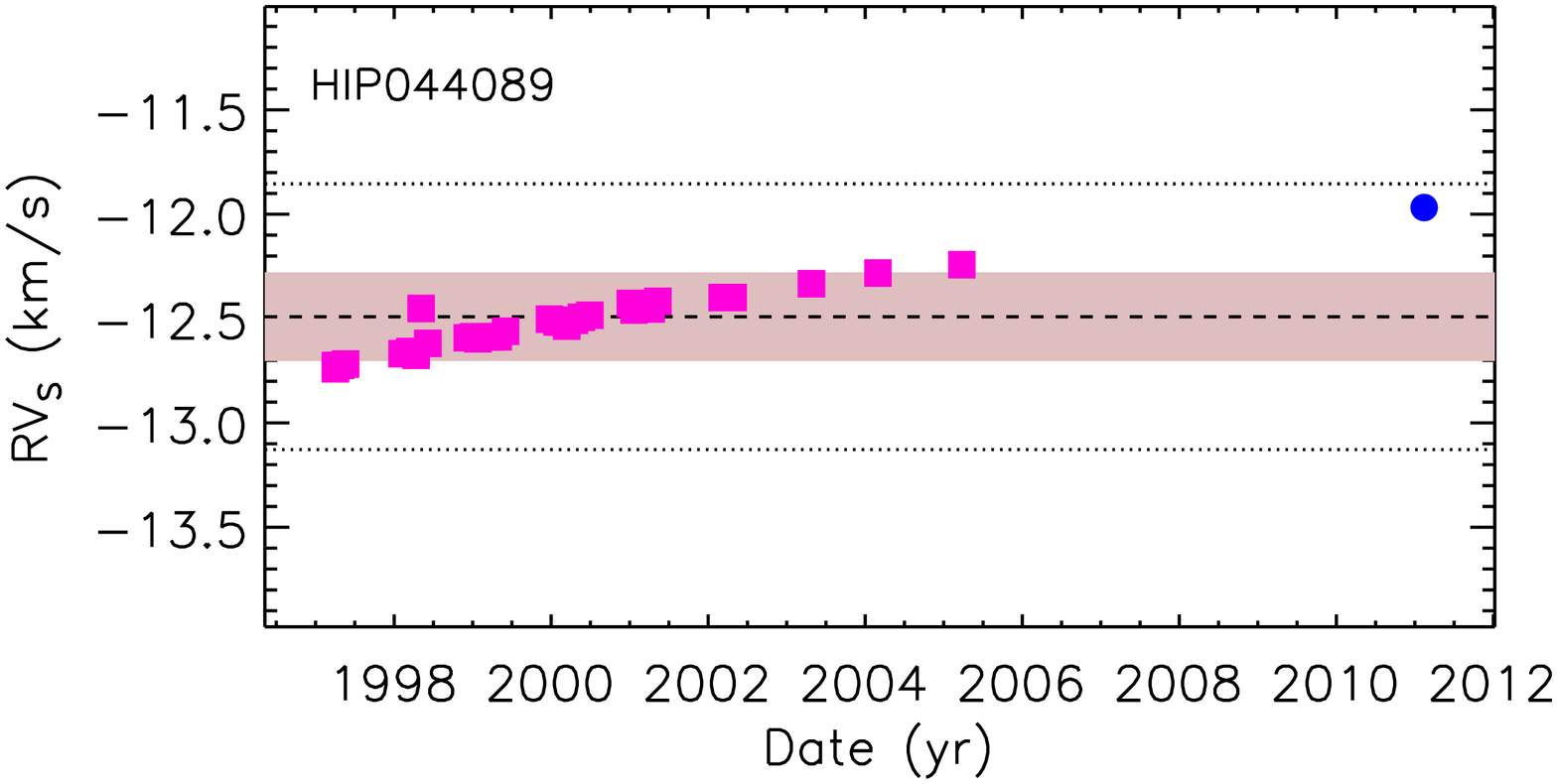}
\includegraphics[width=0.5\textwidth]{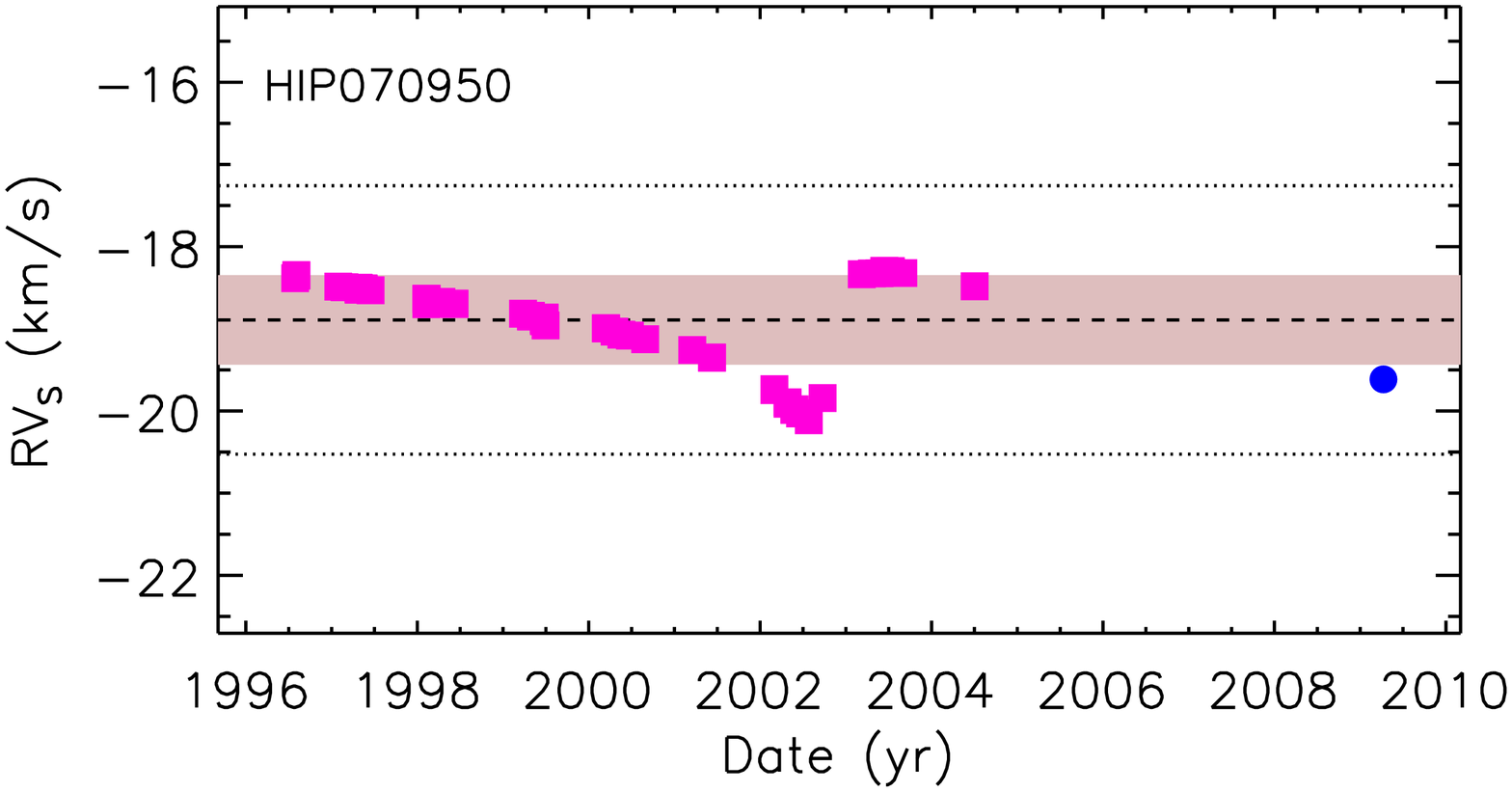}\includegraphics[width=0.5\textwidth]{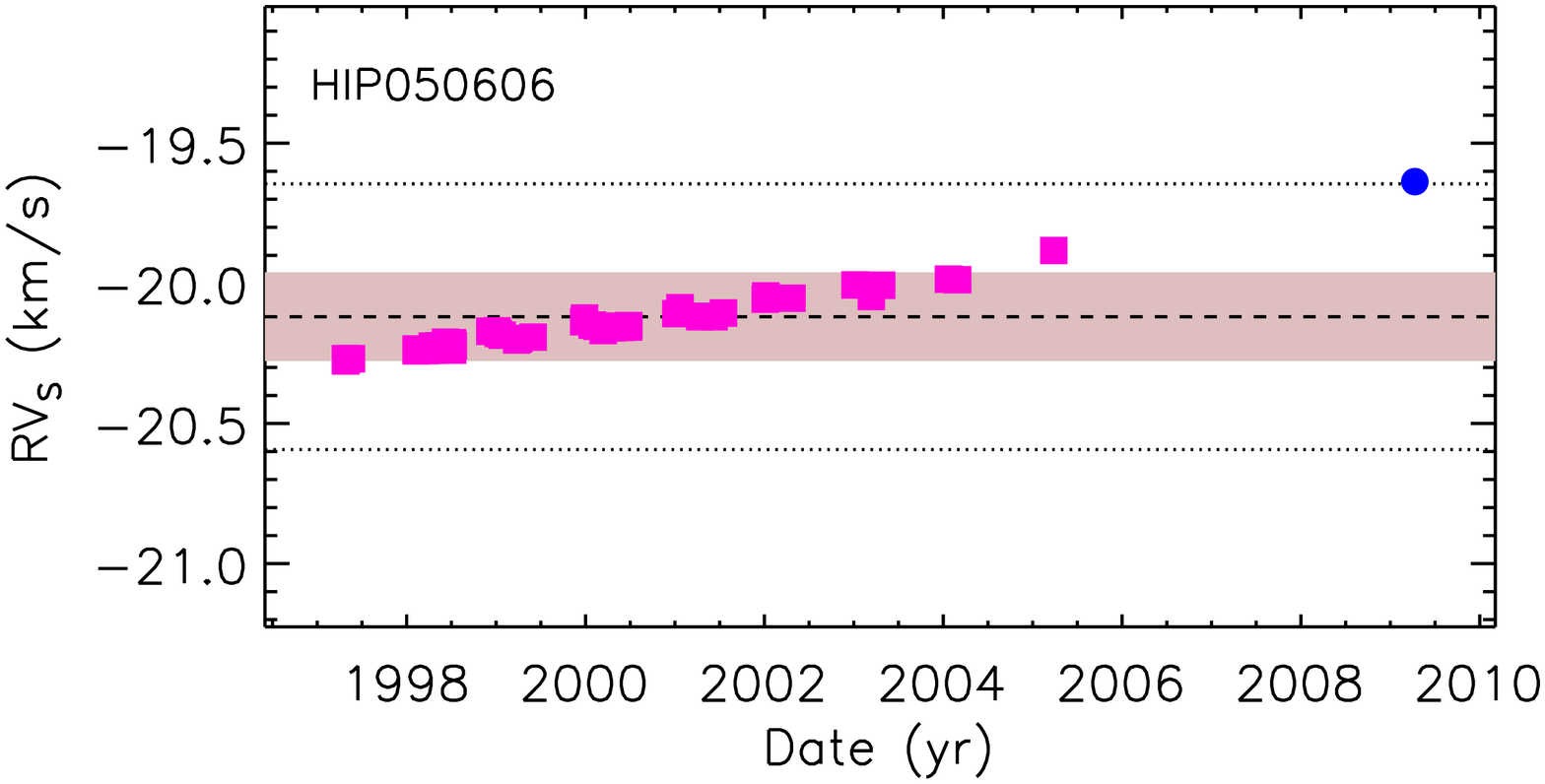}
\includegraphics[width=0.5\textwidth]{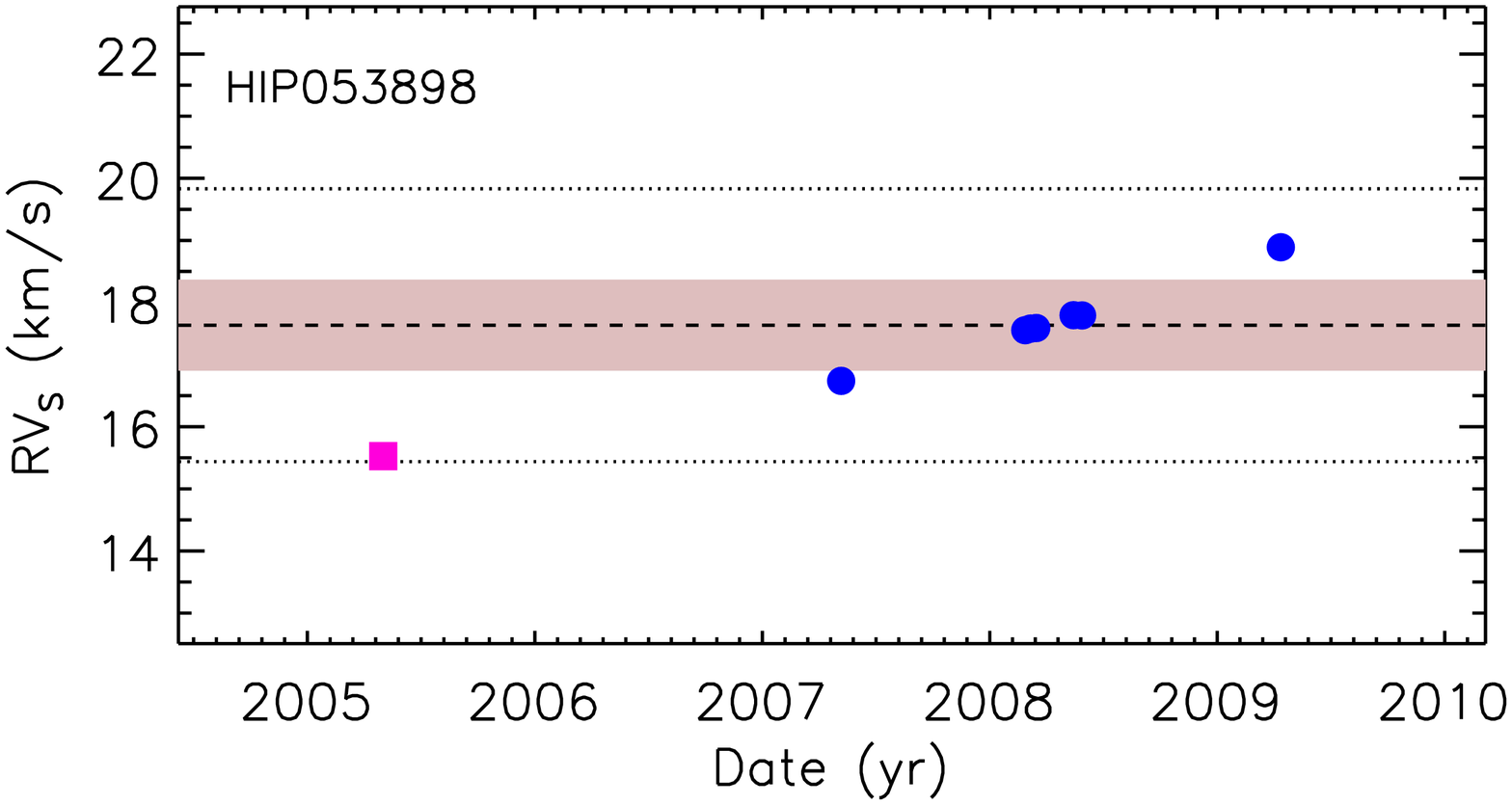}\includegraphics[width=0.5\textwidth]{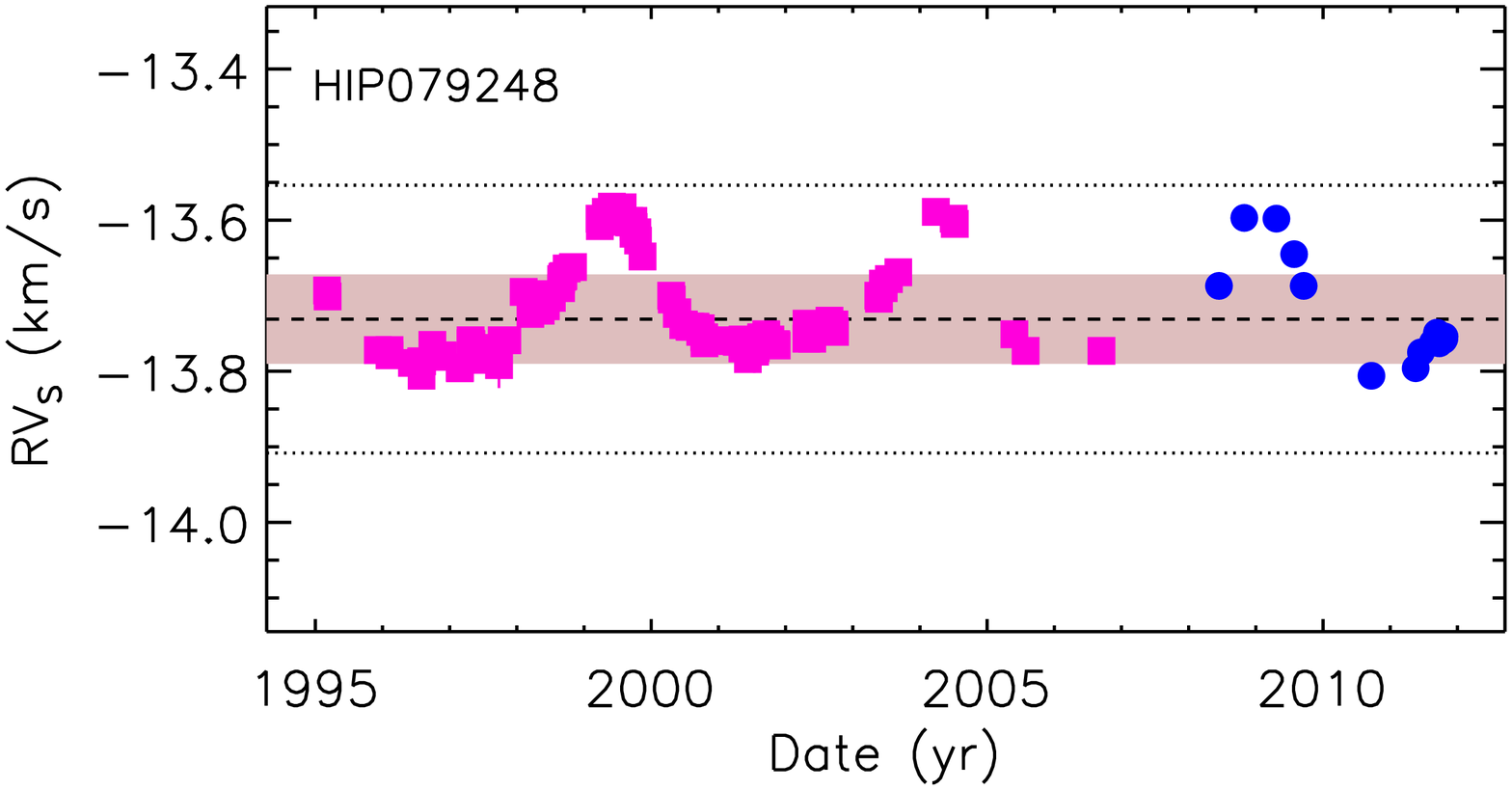}
\includegraphics[width=0.5\textwidth]{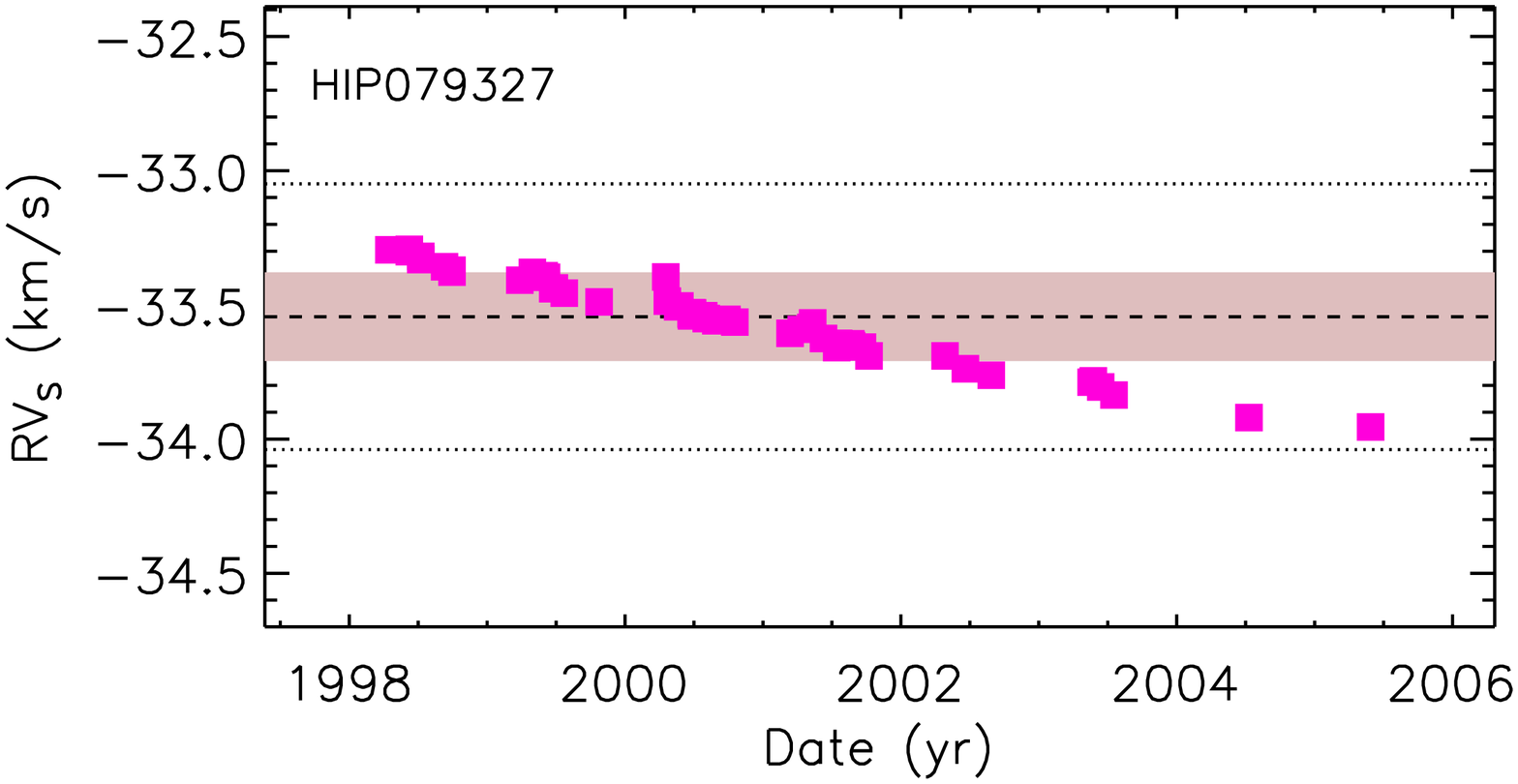}\includegraphics[width=0.5\textwidth]{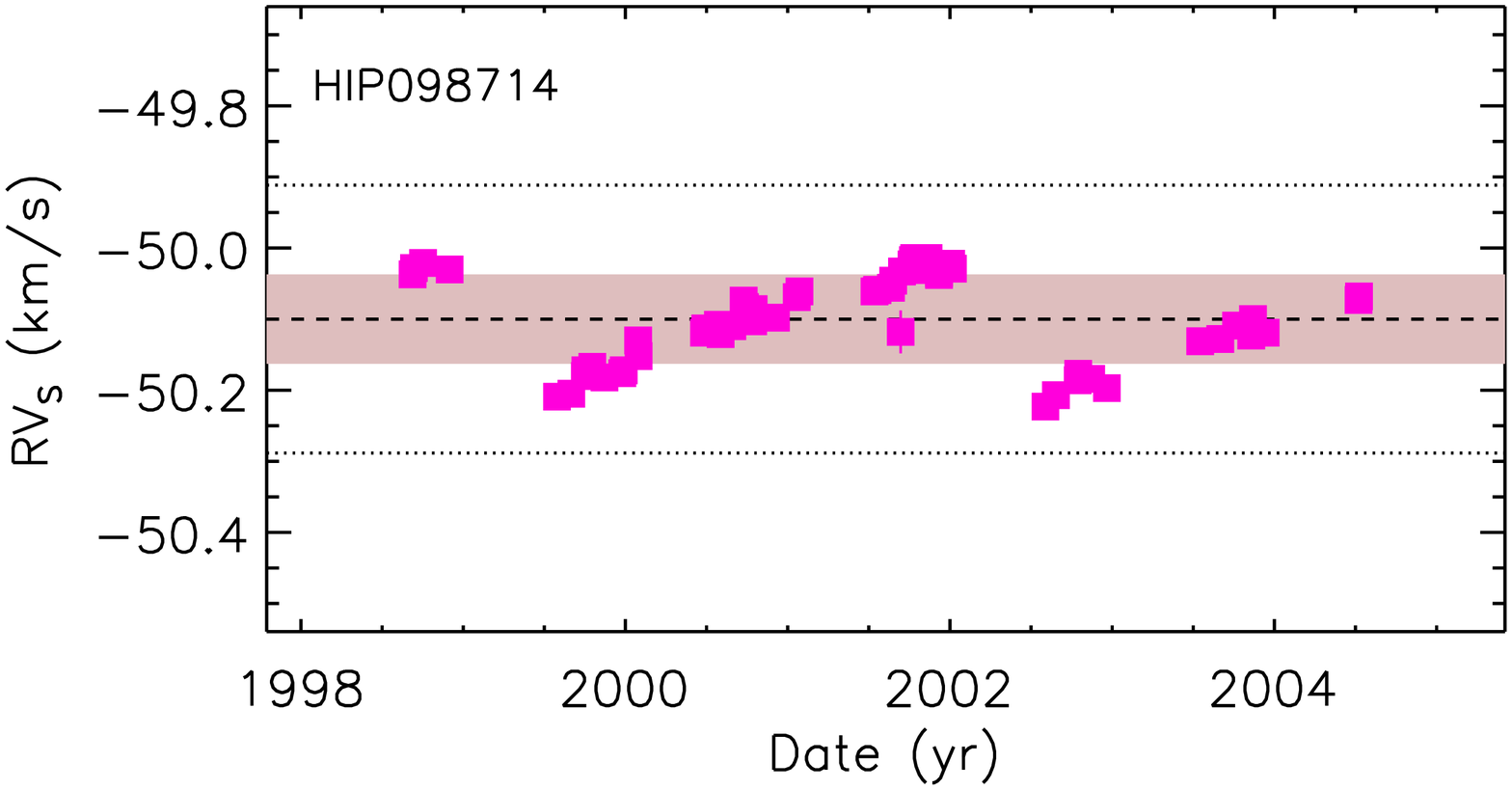}
\caption{Trend in RV measurements for some stars of the catalogue. Symbols as in Fig.~\ref{f:stable_stars}}.
\label{f:variable_stars}
\end{center}
\end{figure*}

Several stars showing a remarkable trend in their  RV are shown in Fig.~\ref{f:variable_stars}. The presence of a low-mass companion likely explains these drifts.  The oscillation of HIP017960 is due to a Jupiter like exoplanet already reported by \citet{boi12}. HIP079248 is also known to harbour two exoplanets \citep{goz06}.  A Keplerian orbit with a period of 1161 $\pm$ 45 days was fitted on HIP098714 measurements by \citet{siv04}, revealing the presence of a planetary companion. Its semi-amplitude of  95 $\pm$ 17 \ms\ is still low enough to make this star suitable for the RVS calibrations. Interestingly, HIP041844 and HIP044089 are part of the Nidever et al. stable stars. The drift of these two stars is quite slow, which might explain why Nidever et al. have not been able to detect it within the time range of their observations. This demonstrates that it is very important to have observations over a sufficient time base to detect such variations. 
However, the number of binary stars remaining in our sample is expected to be very low. The selection criteria described in \citet{cri10} and recalled in Sect.~\ref{sec:candidates} eliminated all known double stars and variable stars (photometrically and spectroscopically), as well as those with fainter neighbours. HIPPARCOS efficiently detected double stars with separations of 0.1\arcsec\ to 10\arcsec, and magnitude differences lower than 4 magnitudes (see Fig. 3.2.106 available in the HIPPARCOS Catalogue\footnote{\tiny 
{on-line at \\
\url{www.rssd.esa.int/index.php?project=HIPPARCOS&page=Overview}}} volume 1 ). Such stars are labelled in HIPPARCOS and were not selected by \citet{cri10}. This guarantees that there is no spectroscopic binaries with short-term variations in our sample, because  the candidates are very nearby stars, all closer than 150pc except some giants. Only a few wide systems with a faint secondary (M star or white dwarf) and a rather long period may therefore remain in the sample. For those, the second round of observations during the Gaia mission will be decisive in constraining their long-term behaviour. With new observations to be done after 2014, we will extend the time baselines by several years (current median epoch of our observations is 2006.9), and we will have at least three measurements per star (67\% of the stars have already 3 measurements or more), well distributed in time and with a much higher precision than the threshold of 300 \ms. This will guarantee that we cannot miss any long period binary with large amplitude that would not be suitable for the RVS calibrations.

\begin{figure}[h!]
\begin{center}
 \includegraphics[width=0.5\textwidth]{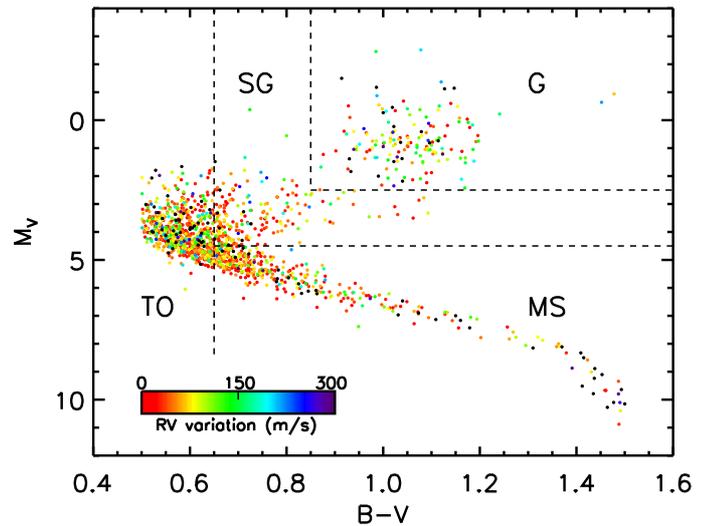}
\caption{HR diagram of the 1420 candidates with a colour code corresponding to their stability level 3$\sigma_{\rm RV_ S}$. The less stable stars, with $3\sigma_{\rm RV_S} > 300$ \ms, are indicated as black points. Those are not suitable for calibrating the RVS instrument. The HR diagram is divided into four parts corresponding roughly to turn-off, main sequence, subgiants, and giants where median RV scatter, and the fractions of stable and variable stars have been evaluated (Table~\ref{t:HR_counts}).}
\label{f:HR_var}
\end{center}
\end{figure}

It is interesting to investigate whether the most stable stars, with $\sigma_{\rm RV_S} < 30$ \ms, or on the contrary the less stable stars, with $\sigma_{\rm RV_S} > 100$ \ms, display in peculiar locations of the HR diagram. Figure~\ref{f:HR_var} shows the HR diagram of the 1420 candidates with a colour code corresponding to their stability level $3\sigma_{\rm RV_{\rm S}}$. The median $\sigma_{\rm RV_{\rm S}}$ and the fraction of  stable and non-stable stars are evaluated into four parts of HR diagram, as indicated in  Fig.~\ref{f:HR_var} : main sequence ($M_V > 4.5,  B-V > 0.65$), turn-off  ($B-V \leq 0.65$),  giants ($M_V \leq 2.5,  B-V > 0.85$), and sub-giants (the rest of the sample). Results of the counts are given in Table~\ref{t:HR_counts}. They clearly show that giants are less stable on average than dwarfs and subgiants, confirming previous findings \citep{dol05}. Not only is the presence of a stellar or substellar companion orbiting around the
giant star suspected to cause the RV variations, but also the possible rotational
modulation of surface inhomogeneities or pulsations.

\begin{table}[h]
  \centering 
  \caption{Percentage of stable ($\sigma_{\rm RV_S} < 30$ \ms) and variable ($\sigma_{\rm RV_S} > 100$ \ms) stars among our 1420 selected candidates, in four parts of the HR diagram.}
  \label{t:HR_counts}
\begin{tabular}{|l|c|c|r|r|}
\hline
HR part & N & median $\sigma_{\rm RV_{\rm S}}$ & stable  &  variable \\
              & & \ms & \%  & \% \\
\hline
main sequence & 397 & 17.0 & 76 & 9\\ 
turn-off & 662 & 18.3 & 74 & 6\\ 
subgiants & 199 & 17.7  & 75 & 5\\ 
giants &154  & 34.6  & 46  & 11\\ 
 \hline
\end{tabular}
\end{table}

\section{Next steps}
\label{s:next}
 There are many physical mechanisms that affect spectroscopic radial velocities \citep{lin03}.
For instance, the convective shifts caused by motions in 
stellar atmospheres depend on stellar lines and on the temperature and gravity. They can 
reach $3$ \kms\ for a Ca{\sc ii} line in an F dwarf, and $-0.4$ \kms\ for an Fe{\sc i} line 
in a K giant \citep{chi11}. Other astrophysical processes may
affect the spectroscopic RV of a star, such as its rotation and activity, the granulation, or the presence 
of low-mass companions.  

 One of the objectives of our ongoing project will be to provide an accurate kinematical RV for each of the 1420 stars in the sample. This  
will be done by deriving the gravitational redshift from  atmospheric parameters and the 
convective shift corrections from three-dimensional hydrodynamical 
model atmospheres. Another  step will be to determine the zero point of the catalogue by comparing the RV measurements of a selection of asteroids with their 
 kinematical velocities as derived from celestial mechanics. 
 Finally, we will continue the observations of the 1420 candidate stars during the Gaia mission 
 in order to validate the long-term stability of their RV. All those steps are essential for calibrating the measurements that will 
 be performed with the Gaia-RVS.

\section{Conclusion}
We have presented the pre-launch version of the catalogue of RV standard stars for Gaia, 
assembled thanks to a long-term observing programme started in 2006 on several spectrographs and with archived measurements. The precision of this catalogue is at the same level as that of \citet{nid02}, $\sim$ 33 \ms, for FGK stars ($0.5 \lesssim B-V \lesssim 1.2$), showing that the RV measurements from the different instruments have been properly combined.
The vast majority of the 1420 selected candidates are found to be stable at the 300 \ms\ level, 
which makes them suitable for calibrating the RVS instrument. Their long-term stability 
will have to be confirmed with new ground-based observations during the Gaia operations. 

 The stable stars presented here have characteristics that make them very useful as standards for many projects other than Gaia. They have a good coverage of important parameters, such as sky distribution, apparent magnitude, spectral type, and luminosity class. They can be used to intercompare intruments and calibrations of zero point of spectroscopic observations. The stars and measurements of this list and those of all other studies on RV-calibrators will soon be discussed within the task group
'radial-velocity standard stars' of the IAU Commission 30. 

\begin{acknowledgements}
 We are indebted to AS-Gaia, PNPS, and PNCG  for their financial support of the observing campaigns 
 and help in this project. We thank Fr\'ed\'eric Arenou for valuable discussions about the contamination of our sample by binary stars.
We warmly thank Sergio Ilovaisky, Maxime Marmier, and Dominique Naef for helping us retrieve relevant data in the OHP and Geneva archives. 
Many thanks also go to staff of the observatories who made observations, and to Lionel Veltz who also contributed to them. We also thank the staff 
maintaining the public archives of ready-to-use spectra at OHP and ESO.   LC acknowledges a financial support from CNES. 
\end{acknowledgements}

\end{document}